\documentclass[aps,prl,twocolumn,superscriptaddress,longbibliography,nobibnotes]{revtex4-2}

\usepackage{amsfonts}
\usepackage{amssymb}
\usepackage{amsmath}
\usepackage{upgreek}
\usepackage{color}
\usepackage{lineno}
\usepackage{graphicx}
\usepackage{hyperref}
\hypersetup{colorlinks,breaklinks,
            urlcolor=[rgb]{0,0,0.72},
            linkcolor=[rgb]{0,0,0.72},
            citecolor=[rgb]{0,0,0.72},
            filecolor=[rgb]{0,0,0.72}}
\usepackage[T1]{fontenc}

\newcommand{\beginsupplement} {
    \setcounter{table}{0}
    \renewcommand{\thetable}{S\arabic{table}}
    \setcounter{figure}{0}
    \renewcommand{\thefigure}{S\arabic{figure}}
    \setcounter{equation}{0}
    \renewcommand{\theequation}{S\arabic{equation}}
    \setcounter{section}{0}
    \setcounter{secnumdepth}{2}
}

\DeclareMathOperator{\arccot}{arccot}

\makeatletter
\def\@ssect@ltx#1#2#3#4#5#6[#7]#8{%
  \def\H@svsec{\phantomsection}%
  \@tempskipa #5\relax
  \@ifdim{\@tempskipa>\z@}{%
    \begingroup
      \interlinepenalty \@M
      #6{%
       \@ifundefined{@hangfroms@#1}{\@hang@froms}{\csname @hangfroms@#1\endcsname}%
       {\hskip#3\relax\H@svsec}{#8}%
      }%
      \@@par
    \endgroup
    \@ifundefined{#1smark}{\@gobble}{\csname #1smark\endcsname}{#7}%
  }{%
    \def\@svsechd{%
      #6{%
       \@ifundefined{@runin@tos@#1}{\@runin@tos}{\csname @runin@tos@#1\endcsname}%
       {\hskip#3\relax\H@svsec}{#8}%
      }%
      \@ifundefined{#1smark}{\@gobble}{\csname #1smark\endcsname}{#7}%
      \addcontentsline{toc}{#1}{\protect\numberline{}#8}%
    }%
  }%
  \@xsect{#5}%
}%
\makeatother

\newcommand{\mytitle}{Delamination-assisted ultrafast wrinkle formation in a freestanding film}

\newcommand{\MIT}{Massachusetts Institute of Technology, Department of Physics, Cambridge, Massachusetts 02139, USA.}
\newcommand{\StanforAP}{Department of Applied Physics, Stanford University, Stanford, California 94305, USA.}
\newcommand{\UCB}{University of California at Berkeley, Department of Chemistry, Berkeley, California 94720, USA}
\newcommand{\SIMES}{SIMES, SLAC National Accelerator Laboratory, Menlo Park, California 94025, USA.}
\newcommand{\UCLA}{Department of Physics and Astronomy, University of California, Los Angeles, California 90095, USA. }
\newcommand{\UHous}{Department of Physics and Texas Center for Superconductivity, University of Houston, Houston, Texas 77204, USA. }
\newcommand{\DESY}{Center for Free-Electron Laser Science, DESY, Notkestra{\ss}e 85, 22607 Hamburg, Germany.}
\newcommand{\UCD}{Department of Materials Science and Engineering, University of California, Davis, California 95616, USA. }
\newcommand{\USTC}{School of Microelectronics, University of Science and Technology of China, Hefei, Anhui 230026, China }

\begin{document}
\title{Delamination-assisted ultrafast wrinkle formation in a freestanding film}

\author{Yifan~Su}
\thanks{These authors contributed equally to this work: Y.S. and A.Z.}
\affiliation{\MIT}
\author{Alfred~Zong}
\thanks{These authors contributed equally to this work: Y.S. and A.Z.}
\affiliation{\UCB}
\affiliation{\MIT}
\author{Anshul~Kogar}
\thanks{Present address: \UCLA}
\affiliation{\MIT}
\author{Di~Lu}
\thanks{Present address: \USTC}
\affiliation{\StanforAP}
\affiliation{\SIMES}
\author{Seung~Sae~Hong}
\thanks{Present address: \UCD}
\affiliation{\StanforAP}
\affiliation{\SIMES}
\author{Byron~Freelon}
\thanks{Present address: \UHous}
\affiliation{\MIT}
\author{Timm~Rohwer}
\thanks{Present address: \DESY}
\affiliation{\MIT}
\author{Harold~Y.~Hwang}
\affiliation{\StanforAP}
\affiliation{\SIMES}
\author{Nuh~Gedik}
\email[Correspondence to: ]{gedik@mit.edu}
\affiliation{\MIT}

\date{\today}

\begin{abstract}
The ability to synthesize and control two-dimensional (2D) crystals creates numerous opportunities for studying emergent states of matter and their novel functionalities. Freestanding films provide a particularly versatile platform for materials engineering in 2D thanks to additional structural motifs not found in films adhered to a flat substrate. A ubiquitous example is wrinkles, yet little is known about how they can develop over as fast as a few picoseconds due to a lack of experimental probes to visualize their dynamics in real time at the nanoscopic scale. Here, we use ultrafast electron diffraction to directly observe light-activated wrinkling formation in a freestanding La$_{2/3}$Ca$_{1/3}$MnO$_3$ film. Via a ``lock-in'' analysis of a single mode of oscillation in the diffraction peak position, intensity, and width, we are able to quantitatively reconstruct how wrinkles develop at the picosecond timescale. Contrary to the common assumption of a fixed boundary condition for freestanding films, we found that wrinkle development is associated with ultrafast delamination at the contact point between the film boundary and the underlying substrate, avoiding a strain build-up in the film. Not only does our work provide a generic protocol to quantify wrinkling dynamics in freestanding films, it also highlights the importance of film-substrate interaction for determining the properties of freestanding structures.
\end{abstract}

\maketitle

\begin{figure*}[tb!]
\centering
\includegraphics[width=1\textwidth]{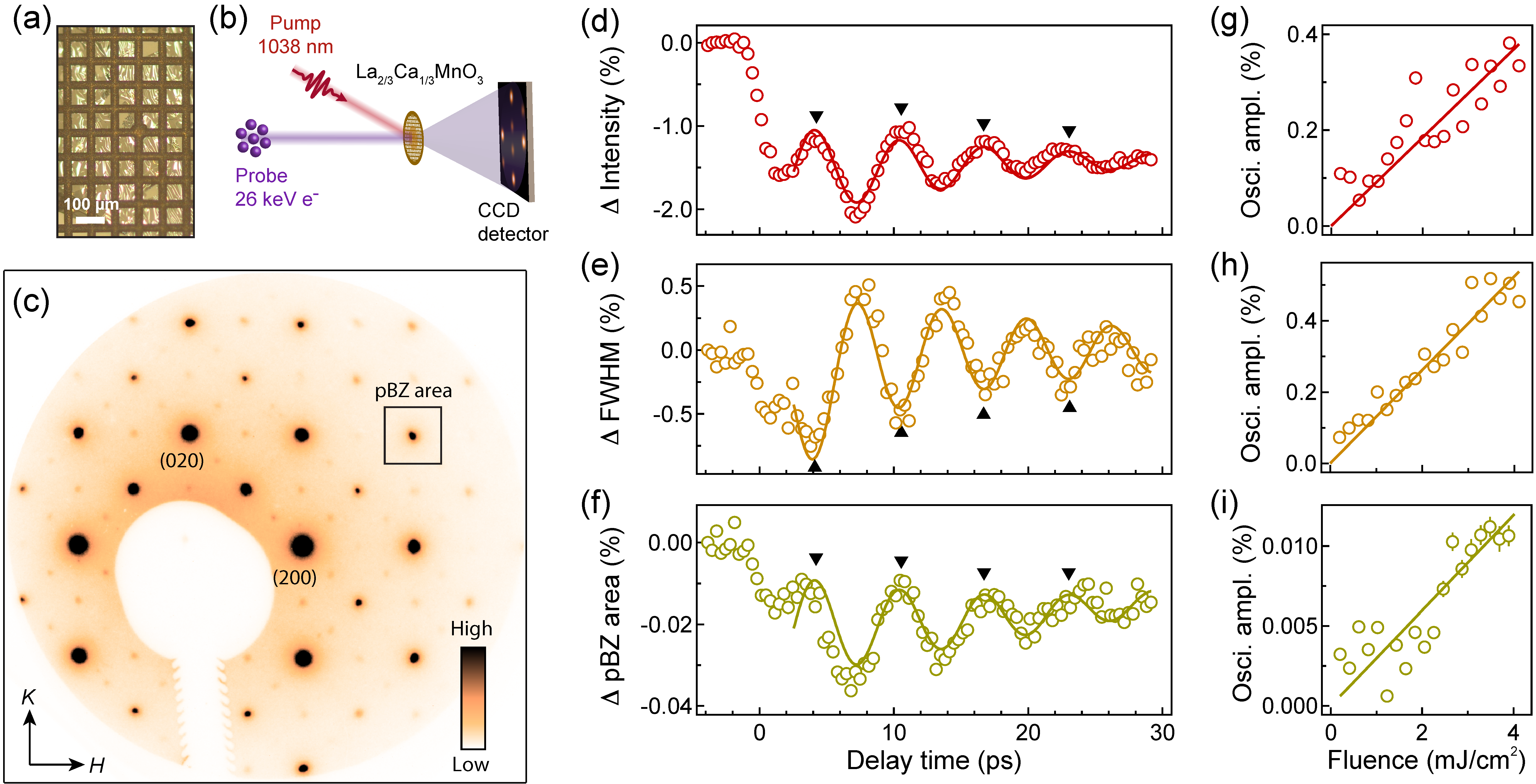}
    \caption{Observation of coherent wrinkling motion in freestanding thin films. (a)~Optical micrograph of a 20-nm-thick, freestanding LCMO film on copper TEM grids. (b)~Schematic of the ultrafast electron diffraction experiment. (c)~Equilibrium electron diffraction pattern of a LCMO thin film at room temperature. The Miller indices of two Bragg peaks are labeled using the pseudo-cubic unit cell. The projected Brillouin zone (pBZ) is marked by a black box. (d)--(f)~Time evolution of the percentage change in peak intensity (d), FWHM (e), and pBZ area (f). Solid curves are sinusoidal fits with an exponential decay \cite{Remark1}. All three quantities show oscillations with the same 0.16~THz frequency. Relative phase relations of the oscillation are highlighted by the black triangles. The incident pump laser fluence was 4.1~mJ/cm$^2$. (g)--(i)~Oscillation amplitudes of the three quantities in (d)--(f) as a function of incident pump pulse fluence. Lines are linear fits.}
\label{fig:main}
\end{figure*}

Recent advances in graphene research have unveiled a lattice degree of freedom that is unique to a 2D freestanding structure: \emph{wrinkles}. They manifest as undulations in thin crystals due to lattice instability, strain relaxation, or thermal expansion mismatch \cite{Nikravesh2020}. For laterally extended films, wrinkles are readily visible in optical micrographs [Fig.~\ref{fig:main}(a)], where alternating bright and dark fringes result from the undulating film height. From the viewpoint of materials engineering, designing specific wrinkles in a graphene sheet can modify its surface wettability and electrochemical properties \cite{Chen2016}. Wrinkle-induced strain gradient can also lead to flexoelectricity and flexomagnetism --- two novel pathways towards designing functional materials \cite{Harbola2021,Zubko2013,Lukashev2010}. From the perspective of symmetry, non-uniform strain caused by wrinkles breaks local inversion symmetry in the lattice \cite{LeQuang2021,Kolhatkar2019,Mennel2019}, allowing nominally forbidden nonlinear optical processes to occur. In graphene, such non-uniform strains can further induce a giant pseudo-magnetic field \cite{CastroNeto2009}, which quantizes electron orbitals into Landau levels and gives rise to flat bands and localized electrons \cite{Milovanovic2020}. Thus, wrinkle engineering not only modifies the mechanical properties of freestanding crystals, it also emerges as a frontier in tuning their electronic correlation and optical response.

Despite the word ``freestanding'', the boundary of such films are still in contact with some substrate and only the interior is suspended. A key but often overlooked aspect of freestanding films is hence their interaction with the underlying substrate via dangling bonds or interfacial chemical reactions \cite{Sung2014,Jiang2020}, especially in the presence of wrinkling.  For non-freestanding films adhered to an elastic substrate, it has been shown that a wrinkling-to-delamination transition can take place under substrate compression, where the film can be partially detached from the substrate \cite{Nolte2017, Vella2009}. For freestanding films, however, it is still unclear how wrinkles form or disappear in relation to their interaction of the substrate at the film boundary. 

Due to a lack of probes to visualize the motion of freestanding films, where displacements often occur on the picometer spatial scale and picosecond timescale, a microscopic understanding of wrinkle development remains missing. In this Letter, we use time-resolved electron diffraction to directly access the structural mode associated with wrinkles in a freestanding film. Remarkably, we find that wrinkle dynamics are accompanied by ultrafast delamination of the film at its boundary contact with the substrate. This discovery is enabled by a gigahertz ``lock-in'' measurement \cite{Gerber2017}, where a single structural resonance is excited by a femtosecond laser pulse so that we can precisely measure its effect on the wrinkles free from a background of incoherent signals. Using a simple model of diffraction, we quantitatively reconstruct the film motion, providing a unique perspective on the mechanism of wrinkle formation and strain relief that occur over a few picoseconds.

In this study, we measured a freestanding film of La$_{2/3}$Ca$_{1/3}$MnO$_3$ (LCMO), a functional oxide known for the colossal magnetoresistance effect \cite{Dagotto2001}. While our findings are not specific to this particular compound, it presents as an interesting candidate because strains associated with the wrinkles may give rise to a number of exotic electronic states \cite{Zhang2016,Chen2020,McLeod2020,Hong2020}. The film was prepared by growing a heterostructure of SrTiO$_3$, SrCa$_2$Al$_2$O$_6$, and LCMO using pulsed laser deposition. The sacrificial layer, SrCa$_2$Al$_2$O$_6$, was subsequently etched away by deionized water to yield a freestanding LCMO membrane \cite{Lu2016,Hong2017,Ji2019,Paskiewicz2016,Kum2020,Bakaul2016}, an image of which mounted on standard TEM grids is shown in Fig.~\ref{fig:main}(a). 

The schematic of the ultrafast electron diffraction experiment is illustrated in Fig.~\ref{fig:main}(b), where a 26~keV electron pulse was used to probe the structural dynamics of a 20-nm-thick LCMO film at room temperature following photoexcitation by a 190-fs laser pulse centered at 1038~nm \cite{Remark1}. Figure~\ref{fig:main}(c) shows a typical electron diffraction pattern taken in equilibrium. From this image, we can extract three separate quantities: (i)~the intensity of Bragg peaks, (ii)~the full width at half maximum (FWHM) of the peaks, and (iii)~the projected Brillouin zone (pBZ) area, as defined in Fig.~\ref{fig:main}(c). Ref.~\cite{Remark1} explains how we extracted these quantities by fitting the diffraction peak profile.

Upon photoexcitation, all three observables---peak intensity, peak width, and pBZ area---start to execute a 0.16~THz acoustic oscillation, shown in Fig.~\ref{fig:main}(d)--(f). The oscillation amplitudes are approximately 0.4\% for peak intensity, 0.5\% for FWHM, and 0.01\% for pBZ area, where numbers are quoted relative to respective equilibrium values prior to pump pulse incidence. These percentage changes are at least 10 times smaller compared to previous diffraction studies of coherent acoustic modes in freestanding films \cite{Park20053,Harb2009,Moriena2012,Wei2017,Qian2020}. Thanks to the high signal-to-noise ratio delivered by our instrument \cite{Remark1}, we can clearly resolve these small changes, and for the first time, an oscillation in the peak width is observed [Fig.~\ref{fig:main}(e)]. 

\begin{figure}[tb!]
\centering
\includegraphics[width=0.9\columnwidth]{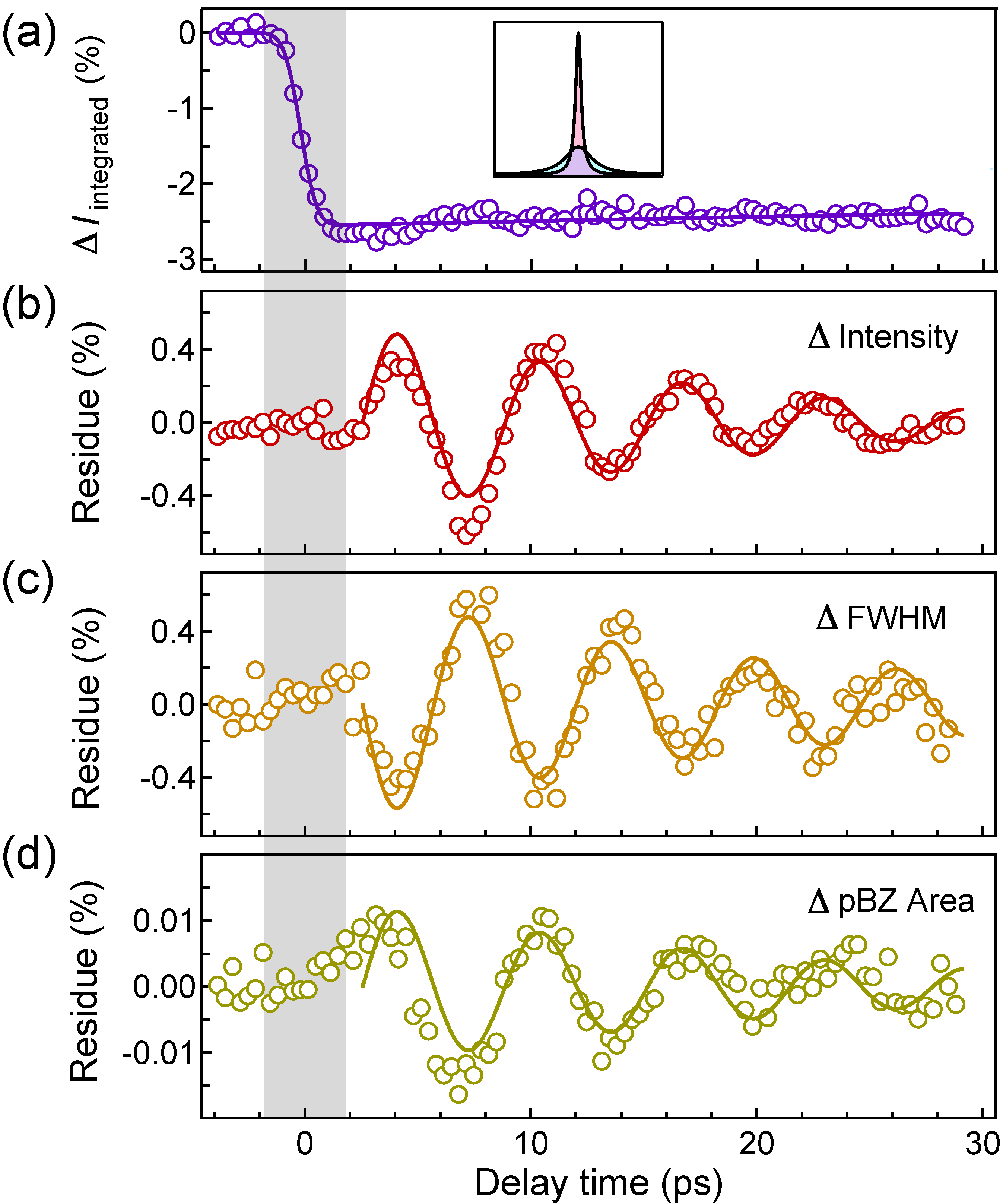}
    \caption{Generation mechanism of the coherent film wrinkling. (a)~Time evolution of the integrated intensity of Bragg peaks, showing a decrease over 2~ps due to the Debye-Waller effect (gray shaded area). No significant oscillation is observed after 2~ps, indicating that the area under the peak profile is conserved (inset). (b)--(d)~Oscillatory components in the time evolution of Bragg peak intensity (b), FWHM (c), and pBZ area (d) after incoherent backgrounds are subtracted from Fig.~\ref{fig:main}(d)--(f). Solid curves are sinusoidal fits with an exponential decay \cite{Remark1}. All three quantities exhibit a delayed oscillation onset by 2~ps relative to the pump pulse arrival.}
\label{fig:res}
\end{figure}

The identical oscillation frequency shared by all three quantities implies a common physical mechanism driving the coherent atomic motion. Importantly, we can solely focus on the oscillatory part of the photoinduced response, shown in Fig.~\ref{fig:res}(b)--(d), and examine what kind of structural movement locks all three quantities into the same mode. To this end, three features of the oscillation stand out. First, they have a delayed onset of approximately 2~ps with respect to the pump pulse incidence [gray shades in Fig.~\ref{fig:res}(b)--(d)]. Second, the peak width oscillation is 180$^\circ$ out-of-phase relative to oscillations in the peak intensity and pBZ area, as highlighted by the black triangles in Fig.~\ref{fig:main}(d)--(f). Lastly, the percentage change of pBZ area is 50-times smaller than that of either peak intensity or width. These characteristics not only provide essential insights into the activation mechanism of the coherent oscillation, they also help us quantitatively reconstruct the real-space film motion that involves, as we show below, wrinkle development accompanied by ultrafast delamination near the boundary of the thin film.

We first analyze the peculiar delay in the oscillation onset time [Fig.~\ref{fig:res}(b)--(d)]. In this discussion, the \emph{peak intensity}, defined by the maximum height of the peak profile, is different from the \emph{integrated intensity}, defined by the area under the profile. Unlike the oscillatory peak intensity after photoexcitation, the integrated intensity of Bragg peaks is dominated by an incoherent suppression with a negligible oscillatory component [Fig.~\ref{fig:res}(a)]. This lack of oscillation accounts for the anti-phase relation between the oscillations in peak intensity and peak width [Fig.~\ref{fig:res}(b)(c)]. In addition, their oscillation amplitudes in terms of percentage changes are similar. This is because their product, a quantity proportional to the integrated intensity, remains constant. The initial decrease of the integrated intensity occurs over 2~ps, highlighted by the gray area in Fig.~\ref{fig:res}(a). This behavior is a manifestation of the Debye-Waller effect, where it takes approximately 2~ps for photoexcited carriers to transfer their excess energy to the lattice within the framework of the two-temperature model \cite{Anisimov1974,Prasankumar2016}. The same 2~ps timescale defines the delayed onset of the coherent oscillations, shown by the gray areas in Fig.~\ref{fig:res}(b)--(d). We thereby conclude that the coherent lattice response originates from a thermoelastic expansion after transient lattice heating by the highly excited electrons, hence leading to the delay \cite{Ruello2015}. As the incident fluence is increased, we observe a linearly proportional increase in the oscillation amplitude for all three quantities [Fig.~\ref{fig:main}(g)--(i)], further confirming the thermoelastic mechanism.

\begin{figure}[tb!]
\centering
\includegraphics[width=\columnwidth]{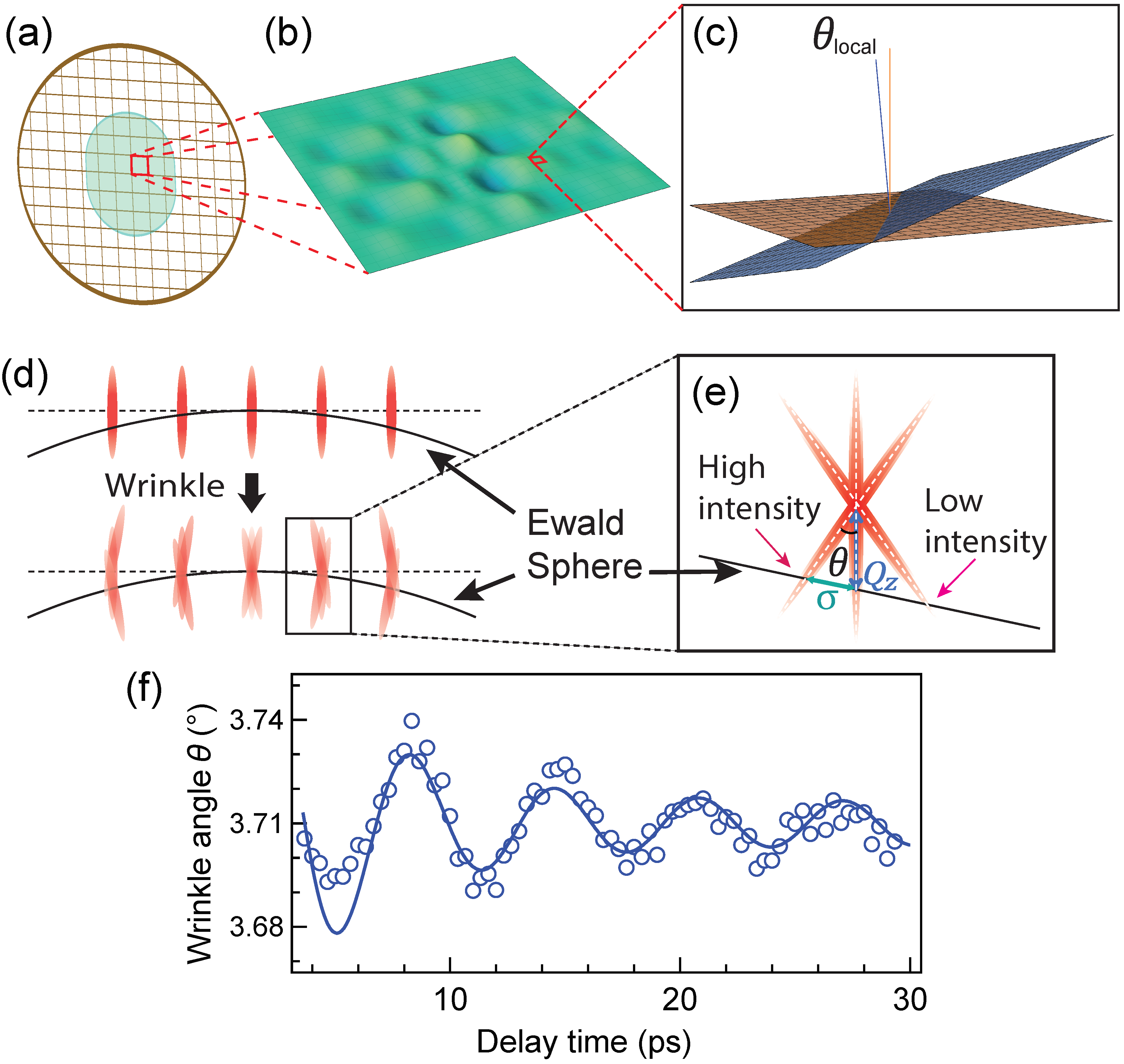}
    \caption{Reconstruction of real-space motion of the wrinkling film. (a)--(c)~Schematic of a wrinkling film on a particular square opening of the copper mesh (b) for a freestanding sample mounted on TEM grids (a). A \emph{local} wrinkle angle, $\theta_\text{local}$, is defined in (c) as the angle between the tilted film plane (blue) and the copper grid plane (orange). (d)~Reciprocal lattice peaks (i.e., rel-rods) before (top) and after (bottom) the development of wrinkles. (e)~An approximate view of an enlarged peak in panel~(d), forming a cone in reciprocal space. The wrinkle angle that characterizes the entire film, $\theta$, is defined as the opening angle of the cone. In the schematic, $\sigma$ is the half-width of the cone, whereas $Q_z$ is the distance from the cone center to the intersection point with the Ewald sphere. Due to an asymmetric intersection between the rel-rod and the Ewald sphere, wrinkling causes a peak shift in momentum space \cite{Remark1}. (f)~Time evolution of the wrinkle angle $\theta$ extracted from Bragg peak width. The solid curve is a sinusoidal fit with an exponential decay \cite{Remark1}.}
\label{fig:extract}
\end{figure}

To reconcile the phase relation between oscillations of the peak width and pBZ area [Fig.~\ref{fig:res}(c)(d)], we need a quantitative model to describe the film motion. We begin by understanding why the peak width oscillates in the first place. In general, a broader diffraction peak indicates a less ordered system, which has been used, for example, to infer the creation of photoinduced defects in the superlattice of charge-density-wave compounds \cite{Vogelgesang2017,Zong2019,Domrose2022}. However, defect generation in the crystal lattice itself requires a much higher absorbed energy density, which leads to a transient melting of solids \cite{Mo2018}. Furthermore, there is no clear mechanism to account for a time-periodic modulation of defect density, rendering this picture inapplicable to our observation. On the other hand, previous measurements on a monolayer MoS$_2$ film suggest that transient formation of wrinkles can also give rise to a broader diffraction peak \cite{Mannebach2015}. While no oscillation of peak width was observed, it is plausible that the degree of wrinkling varies periodically in time due to a photoexcited acoustic mode that modulates the long-wavelength atomic displacement in the film.   

To substantiate the interpretation of a wrinkling oscillation, schematically shown in Fig.~\ref{fig:extract}(a)(b), we define a simple model to quantify this motion and to explore its effect on the different observables. Locally, wrinkling causes a tilt of the film plane [Fig.~\ref{fig:extract}(c)], or equivalently, a rotation of rel-rods in momentum space [Fig.~\ref{fig:extract}(d)]. After we integrate over all randomly-oriented local patches in a macroscopic film illuminated by the electron beam, the superposition of rel-rods for a specific Bragg peak forms two cones that resemble an hour-glass structure [Fig.~\ref{fig:extract}(e)]. We can use a single quantity, henceforth called wrinkle angle $\theta$, to characterize the degree of wrinkling, where $\theta$ is defined as the opening angle of the cones [Fig.~\ref{fig:extract}(e)] \cite{Remark1}. In an electron diffraction experiment, peaks arise at the intersection between the Ewald's sphere and these cones. Therefore, the more wrinkled the sample, the larger $\theta$ would be, and broader Bragg peaks would appear in the diffraction pattern. In Fig.~\ref{fig:extract}(f), we plot the extracted wrinkle angle $\theta$ based on the oscillatory peak width in Fig.~\ref{fig:res}(c). The overall vertical offset of 3.71$^\circ$ comes from the equilibrium wrinkle angle determined from angle-dependent static diffractions \cite{Remark1,Kumar2015}.

In addition to explaining the time evolution of the peak width, the wrinkling motion also accounts for the lack of oscillation in the integrated intensity for a small wrinkle angle $\theta$, as illustrated in Fig.~\ref{fig:extract}(d). When the film is relatively flat, the integrated intensity concentrates in the rel-rods that are perpendicular to the sample plane. When wrinkles develop, rel-rods morph into cones but the total number of intersections with the Ewald's sphere remains the same, thus preserving the integrated intensity. The oscillation of the pBZ area is also consistent with the wrinkling picture. At a qualitative level, the finite curvature of the Ewald's sphere of 26~keV electrons leads to its asymmetric intersection with the reciprocal lattice cones [Fig.~\ref{fig:extract}(e)]. Since the diffraction intensity follows a  $|\text{sinc}\,(TQ_z)|^2$ profile along individual rel-rods in a cone, where $T$ is the film thickness and $Q_z$ is the reciprocal space distance measured from the center of the rel-rod [Fig.~\ref{fig:extract}(e)], such asymmetric intersections lead to a peak shift towards the undiffracted beam when the wrinkle angle increases \cite{Remark1}. Hence, more pronounced wrinkling corresponds to a broader peak and a smaller pBZ area, accounting for their anti-phase relation in the oscillatory signal [Fig.~\ref{fig:res}(c)(d)].

\begin{figure}[tb!]
\centering
\includegraphics[width=1\columnwidth]{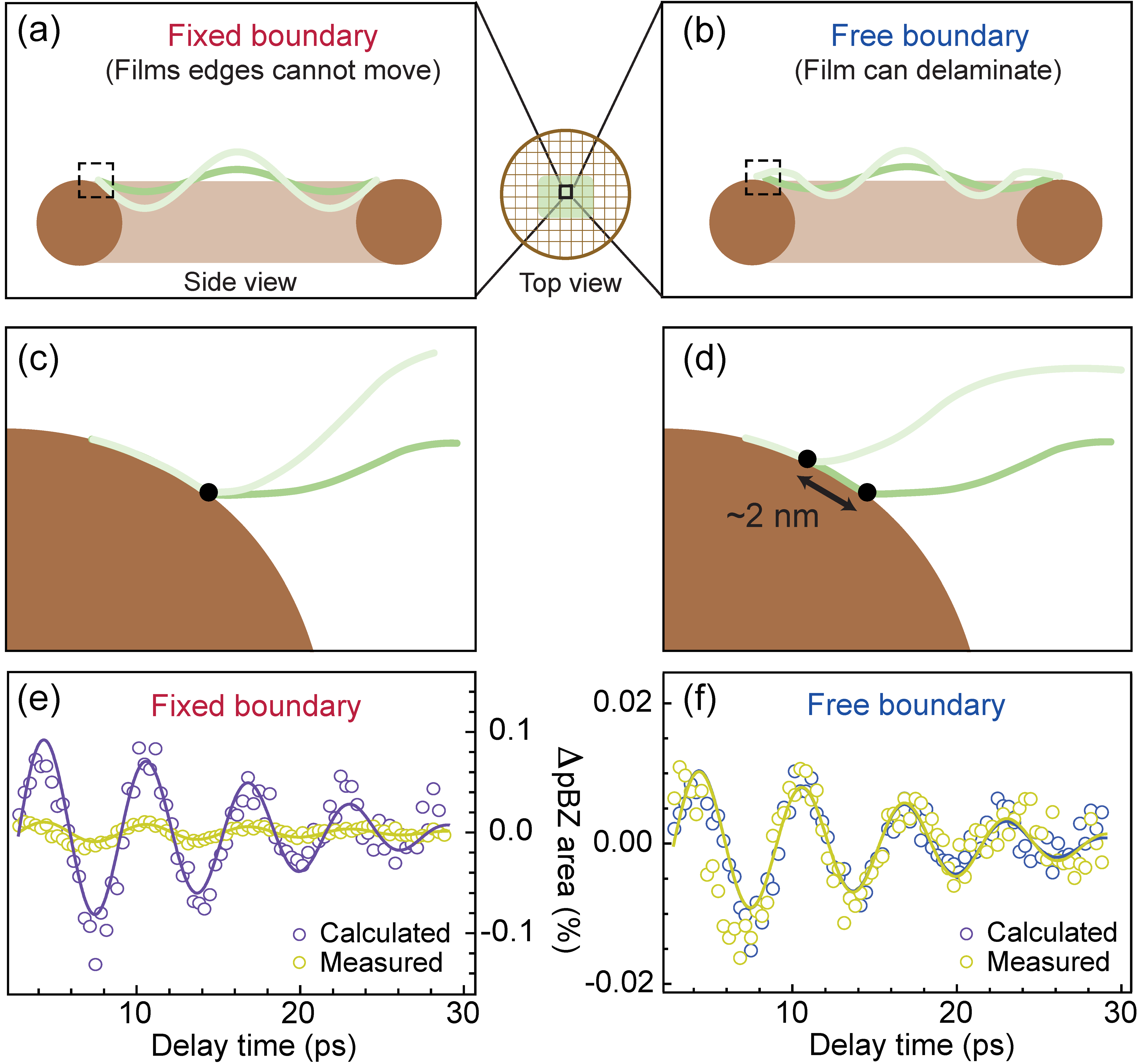}
    \caption{Wrinkle-induced change in the pBZ area for two types of film-substrate interaction. (a)(b)~Schematics of fixed (a) and free (b) boundary conditions at the contact points between the film and the TEM grids, illustrated for a particular opening of the grids. Dark (or light) green curve represents the side view of two wrinkled films with a small (or large) wrinkle angle. (c)(d)~Enlarged view of (a) and (b) respectively, zooming into the contact point between the film and the copper grid. For fixed boundary condition, the film has a fixed contact point with the copper bar. For free boundary condition, the film can delaminate so that the contact point moves during wrinkle formation. The spacing between the light and dark green curves is exaggerated to aid visualization. (e)(f)~Calculated (blue circles) and measured (yellow circles) pBZ area change for the two boundary conditions.  Solid curves are sinusoidal fits with an exponential decay \cite{Remark1}. Note the different vertical scales in the two panels. The calculation was based on the wrinkle angle in Fig.~\ref{fig:extract}(f) extracted from the Bragg peak width.}
\label{fig:bdc}
\end{figure}

While a coherent wrinkling motion justifies the phase relations among oscillations of the peak intensity, peak width, and pBZ area, there remains the stark contrast of a 50-times difference in the oscillation amplitude between the pBZ area compared to the other two quantities [Figs.~\ref{fig:main}(d)--(i) and \ref{fig:res}(b)--(d)]. To rationalize this discrepancy, we need to consider the exact film deformation when wrinkles develop during an oscillation cycle. Specifically, there are two possible motions at the film-substrate contact point, illustrated in Fig.~\ref{fig:bdc}(a)(c) and (b)(d). With a fixed boundary condition [Fig.~\ref{fig:bdc}(a)(c)], wrinkle formation leads to a global stretch of the film surface, introducing strains that lead to an enlarged unit cell and hence a smaller pBZ area. Simple geometric considerations detailed in Ref.~\cite{Remark1} show that the percentage reduction in the pBZ area is in the same order of magnitude as the percentage increase in the wrinkle angle $\theta$ [Fig.~\ref{fig:extract}(f)]. Indeed, the computed pBZ area oscillation based on the $\theta$ oscillation has an amplitude of more than 1.5\% [Fig.~\ref{fig:bdc}(e), purple circles]; however, this value is incongruent with the measured oscillation of the pBZ area [Fig.~\ref{fig:bdc}(e), yellow circles]. The other scenario stipulates a free boundary condition, namely, the film attached to the copper mesh at the boundary can slightly delaminate, effectively having a moving boundary [Fig.~\ref{fig:bdc}(b)(d)]. From the equilibrium configuration to the maximum wrinkle angle reached during an oscillation, the boundary displacement is estimated to be 2~nm, corresponding to an average speed of 1~km/s for the delamination motion \cite{Remark1}. In this case, a larger wrinkle angle does not cause an increase of the global film strain, so the lattice constants remain unchanged. However, the asymmetric intersection between the Ewald's sphere and the reciprocal lattice cones [Fig.~\ref{fig:extract}(e)] still leads to a small but nonzero peak shift, causing a decrease in the pBZ area. The oscillation of this computed pBZ area based on the $\theta$ change is plotted as blue circles in Fig.~\ref{fig:bdc}(f), where the amplitude is in excellent agreement with the tiny oscillation directly derived from the diffraction images (yellow circles).

From the comparative study of these two scenarios, the extremely small oscillation amplitude of the pBZ area points towards an ultrafast delamination near the boundary of the thin film. This picture runs contrary to previous assumption of mechanical pinning at the boundary of a freestanding film \cite{Harb2009,Park2009,Mannebach2015}, suggesting that wrinkle development at the picosecond timescale has negligible contribution from film bulging as depicted in Fig.~\ref{fig:bdc}(a). At the heart of this delamination motion is the tendency for strain relief in a wrinkled film. As shown in Fig.~\ref{fig:strain}(b), the peak-to-peak variation of the maximum strain in the film is merely 0.0002\% during one oscillation cycle. This value is much smaller than the strain arising from a thermoelastic expansion after photoexcitation (Fig.~\ref{fig:thermal_expansion}), indicating that ultrafast wrinkling does not introduce additional strain build-up.

To conclude, by studying three separate observables in time-resolved diffraction patterns -- peak intensity, peak width, and pBZ area -- we reconstruct the wrinkling dynamics of a freestanding film, revealing remarkably rapid delamination of the film from the substrate. Central to this discovery is our ability to resolve extremely small photoinduced changes, allowing for the detection of coherent oscillations in all three quantities. The oscillatory signals have enabled a quantitative description of the ultrafast dynamics unaffected by interference from any incoherent background, akin to a gigahertz ``lock-in'' measurement that is generally applicable to other time-resolved studies \cite{Gerber2017}. These findings provide a new understanding of wrinkle formation at the picosecond timescale that is free from strain build-up. Our work also paves the way for ultrafast manipulation of wrinkles that can in turn modulate electronic, optical, and mechanical properties of nanoscopic quantum materials \cite{CastroNeto2009,Milovanovic2020,Kolhatkar2019,Mennel2019,Lukashev2010,Zubko2013}.

\section*{A\lowercase{cknowledgements}}
We thank Edoardo Baldini, Bai-Qing Lv, Bryan Fichera, Kyoung Hun Oh, Dongsung Choi, Masataka Mogi, Honglie Ning and Minhui Zhu for helpful discussions. We acknowledge support from the U.S. Department of Energy, BES DMSE (UED instrumentation and data acquisition), and from the Gordon and Betty Moore Foundation's EPiQS Initiative grant GBMF9459 (data analysis, manuscript writing). A.Z. acknowledges support from the Miller Institute for Basic Research in Science. T.R. acknowledges the funding from the Humboldt Postdoctoral Fellowship. B.F. would like to thank the MIT Physics Department for its generous support. The work at SLAC/Stanford is supported by the U.S. Department of Energy, BES DMSE under contract number DE-AC02-76SF00515; the Gordon and Betty Moore Foundation’s EPiQS Initiative through grant number GBMF9072 (synthesis equipment).

\providecommand{\noopsort}[1]{}\providecommand{\singleletter}[1]{#1}%
%

\clearpage
\newpage

\onecolumngrid
\begin{center}
\textbf{\large Supplemental Material to ``\mytitle''}
\end{center}\hfill\break
\twocolumngrid

\beginsupplement

\tableofcontents

\makeatletter
\let\toc@pre\relax
\let\toc@post\relax
\makeatother 

\section{E\lowercase{xperimental setup and diffraction peak analyses}}

\subsection{Ultrafast electron diffraction setup}
The 1038~nm (1.19~eV) output of a commercial Yb:KGW regenerative amplifier laser system (PHAROS SP-10-600-PP, Light Conversion),  operating at a base repetition rate 100~kHz and pulse-picked to 5~kHz, was split into pump and probe branches. The pump branch was focused onto the sample for photoexcitation, while the probe branch
was frequency quadrupled to 260~nm (4.77~eV) using two nonlinear crystals and focused onto a gold-coated sapphire in high vacuum ($<3.9\times10^{-9}$~torr) to generate pulsed photoelectron bunches. These electrons were accelerated to 26~kV in a DC field and focused with a solenoid before diffracting from the LCMO film in transmission geometry. Diffracted electrons were incident on an aluminum-coated phosphor screen (P-46), whose luminescence was recorded by a commercial intensified charge-coupled device (iCCD, PI-MAX~II, Princeton Instruments) operating in the shutter mode. More details of the ultrafast electron diffraction setup can be found in Ref.~\cite{ZongThesis}.

\subsection{Peak profile fitting\label{sec:profile_fitting}}

Each peak in the diffraction image is integrated in one dimension and fitted to a Lorentzian profile to determine the peak intensity, width, and position. The chosen line profile follows the Lorentzian profile of the raw electron beam and a linear background is included in the fitting procedure to account for diffuse scattering signals. Each fitting is executed in two perpendicular directions; the arithmetic and geometric means are taken to determine the average intensity and FWHM, respectively \cite{Vogelgesang2018}. This averaging procedure is justified by the absence of in-plane anisotropy other than imperfectness of the raw electron beam. The FWHM of the raw electron beam is subtracted in quadrature to retrieve the intrinsic peak width due to the sample morphology \cite{Mannebach2015,Zong2019}. The fitted peak positions are used to calculate the projected reciprocal lattice constants and thus the projected BZ area, as detailed in the next paragraph. 

\subsection{Global least squares fitting for projected Brillouin zone area extraction\label{sec:BZ_area_fitting}}

A large number of peaks in an electron diffraction image allow us to accurately determine changes in the projected lattice constant in the detector plane. This is accomplished by a weighted regression procedure that takes into consideration all peak positions and their associated uncertainties obtained from the peak profile fittings. Specifically, let $\vec{b}_1$ and  $\vec{b}_2$ denote the projections of the reciprocal lattice unit vectors, and let $ \vec{r}_0 = (x_0,y_0)$ represent the coordinates of the $(000)$ peak in the diffraction image. Given the fitted location of the $i$-th peak $\vec{r}_{i=1,\dots,n}=(x_i,y_i)$ with in-plane Miller indices $\vec{m}_i=(H_i,K_i)$, we have \cite{ZongThesis}:
\begin{align}
    \begin{bmatrix}
    \vec{r}_0\\
    \vec{b}_1\\
    \vec{b}_2
    \end{bmatrix}
    &= \left(M^TWM\right)^{-1}M^TW
    \begin{bmatrix}
     \vec{r}_1\\
    \vdots\\
    \vec{r}_n
    \end{bmatrix}
    , \text{ where}\label{eq:global_fit}\\ 
    M &=
    \begin{bmatrix}
        1 & \vec{m}_1\\
        \vdots & \vdots \\
        1 & \vec{m}_n
    \end{bmatrix}.
\end{align}
Here, $W$ is a diagonal matrix, whose diagonal entries correspond to the reciprocal of the position uncertainty in the $i$-th peak. Solving for $\vec{b}_1$ and  $\vec{b}_2$ from Eq.~\eqref{eq:global_fit} readily yields the value of pBZ area as $\left|\vec{b}_1\times\vec{b}_2\right|$. 

\subsection{Time trace fitting}

In order to extract the coherent oscillation from the total photoinduced response, we need to subtract the incoherent response. For all three quantities -- peak intensity, width, and pBZ area -- the raw data is fitted to an exponential function that models the long-term, incoherent relaxation dynamics multiplied by an error function that models the initial system response. The entire fitting function is convolved with a Gaussian function that models the temporal resolution of the instrument. The fitting function takes the following mathematical form \cite{Zong2019}:
\begin{align}
\Delta f(t)=\Bigg[\frac{1}{2}\left(1+\text{Erf}\left(\frac{2\sqrt{2}(t-t_0)}{w}\right)\right)\notag\\
\cdot(I_\infty+I_0e^{-(t-t_0)/\tau})\Bigg]\ast g(w_0,t).\label{eq:fit}
\end{align}
In this model, $w$ represents the intrinsic system response time due to photoexcitation, in our specific case, the Debye-Waller intensity suppression time. $I_0$ represents the maximum system
response. $I_\infty$ denotes the value of $\Delta f$ at
long time delays, when the system reaches a quasi-equilibrium thermal state. $\tau$ is the characteristic relaxation time to the quasi-equilibrium. $t_0$ is associated with the relative arrival time of pump and probe pulses. 

\begin{figure}[tb!]
\centering
    \includegraphics[width=1\columnwidth]{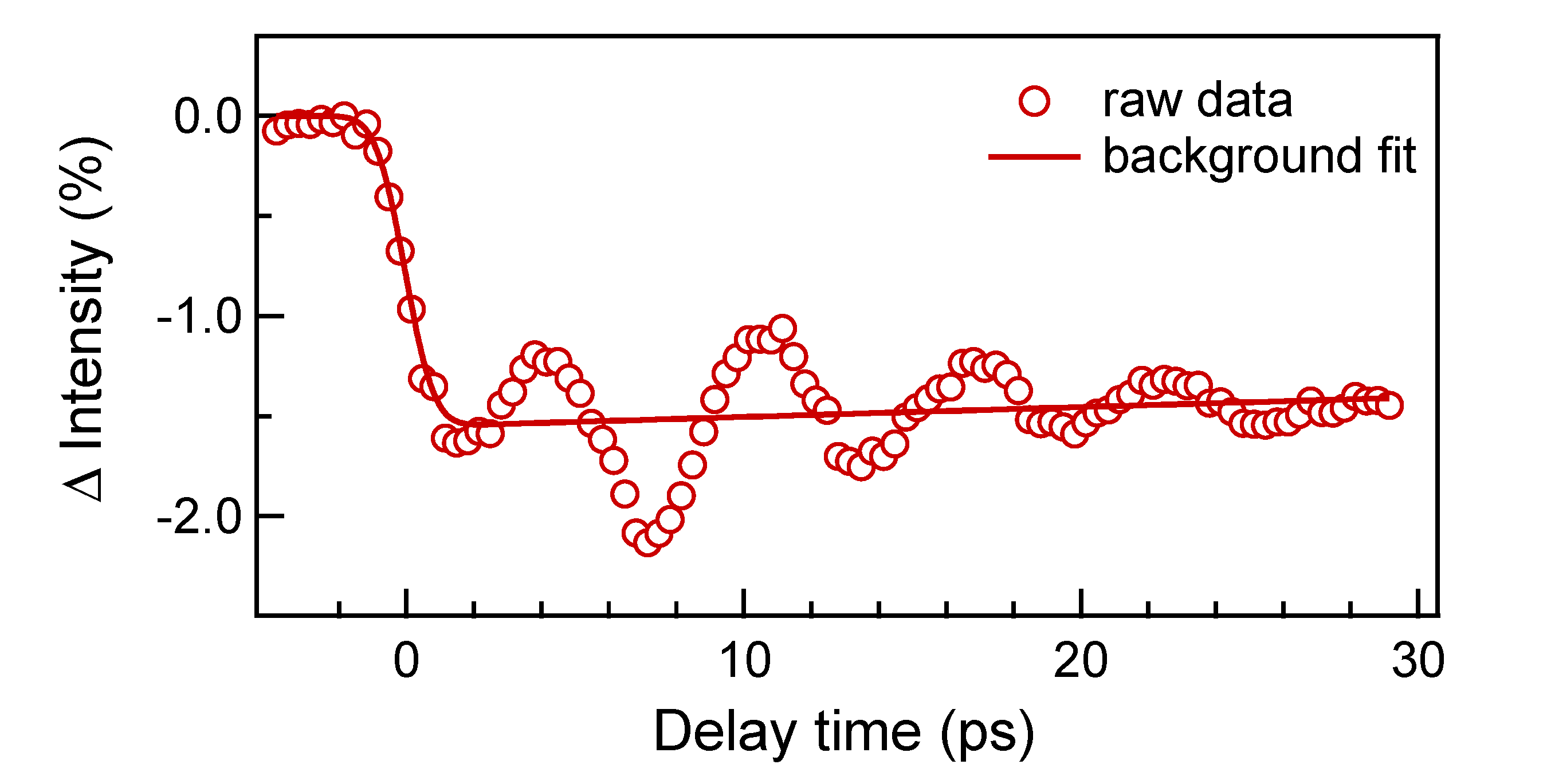}
    \caption{Raw data of photoinduced change in the Bragg peak intensity, reproduced from Fig.~\ref{fig:main}(d). The solid curve is a fit to Eq.~\eqref{eq:fit}, which captures the non-oscillatory background response.}
\label{fig:fit}
\end{figure}

\begin{figure}[tb!]
\centering
    \includegraphics[width=0.7\columnwidth]{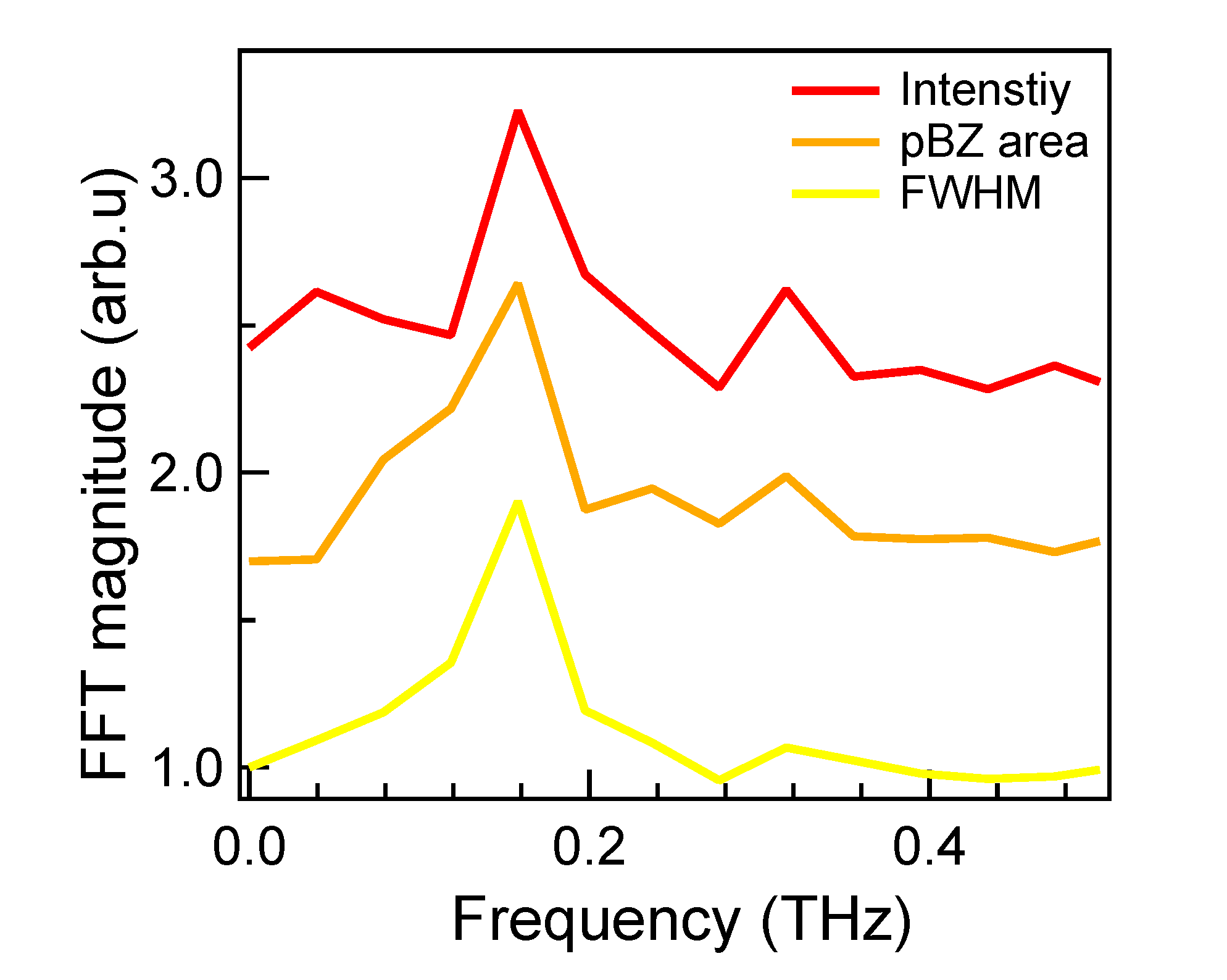}
    \caption{FFT spectra of coherent oscillations in peak intensity, pBZ area, and peak width. The FFT magnitude peaks at 0.16~THz for all three quantities.}
\label{fig:FFT}
\end{figure}

As an example, Fig.~\ref{fig:fit} shows the raw data of photoinduced peak intensity change and the fitted incoherent part to Eq.~\eqref{eq:fit}. The other two quantities share the same fitting procedure. The residue from the curve fitting is then computed and fitted to a cosine function with an exponential decay, as shown in Fig.~\ref{fig:res}(b)--(d). The oscillation frequency is determined to be 0.16~THz both by the cosine fitting and by fast Fourier transform (FFT), shown in Fig.~\ref{fig:FFT}. 

\subsection{Noise level estimates}

The detection of the coherent wrinkling dynamics would not be possible without the extremely low noise floor in our ultrafast electron diffraction data. By calculating the standard deviation of the data points at negative pump-probe delays, we determine the noise levels to be $0.03\%$, $0.1\%$ and $0.003\%$ for peak intensity, peak width, and pBZ area, respectively.

\section{Q\lowercase{uantifying the coherent wrinkling dynamics}}

\subsection{Diffraction peak width and wrinkle angle}

In the main text, we stated that the percentage change in the wrinkle angle can be calculated from the change in the peak width. Here we define a simplified model to connect these two quantities. We first consider a small patch of a thin film. The \emph{local} wrinkle angle, $\theta_\text{local}$, is defined as the local angle between the membrane and the sample plane [Fig.~\ref{fig:extract}(c)]. If we consider the entire film that has a continuous distribution of different $\theta_\text{local}$, the resulting effective wrinkle angle $\theta$ is then defined as the opening angle of the broadened cone formed by rel-rods in reciprocal space [Fig~\ref{fig:extract}(e)]. Similar definition of the wrinkle angle was reported in a previous study of wrinkled monolayer samples using ultrafast electron diffraction \cite{Mannebach2015}. 

Let us consider a locally flat Ewald sphere that intersects the rel-rods [Fig.~\ref{fig:extract}(e)]. The wrinkle angle is given by: 
\begin{align}
    \theta = \frac{\sigma}{Q_z},\label{eq:theta_sigma} 
\end{align}
where $\sigma$ is the half-width of the cone and $Q_z$ is the distance from the cone-center to the intersection point with the Ewald sphere, as indicated in Fig.~\ref{fig:extract}(e). Since $\theta$ is directly proportional to $\sigma$ from Eq.~\eqref{eq:theta_sigma}, we can extract the change in the wrinkle angle $\theta$ from the change in the cone width $\sigma$, which is in turn proportional to the peak width in the diffraction image: 
\begin{align}
    \frac{\Delta\theta}{\theta_0}= \frac{\Delta\sigma}{\sigma_0}=\frac{\Delta \text{FWHM}}{\text{FWHM}_0},\label{eq:theta_FWHM}
\end{align}
where the subscript 0 indicates the equilibrium value.

\subsection{Effect of wrinkling on integrated intensity\label{sec:integrated_intensity}}

The intensity distribution forms a sinc function along the rel-rods \cite{Fultz2002}, which in the thick film limit can be approximated by a Gaussian distribution with a characteristic length $2\pi/T$, where $T=200$~{\AA} is the film thickness. We note that the choice of the distribution function does not have a significant impact on the results.  Under this approximation, the length of the rel-rods is $2\pi/T=0.03\text{~{\AA}}^{-1}$. When a rel-rod tilts by an angle such that the diffraction peak width increases by $\sim1$\% [Fig.~\ref{fig:res}(c)], the point of intersection between the rel-rod and the Ewald sphere moves by a small distance in momentum space, which is on the order of $10^{-6}\text{~{\AA}}^{-1}$. This movement is negligible compared to the length of the rel-rods.

\begin{figure}[tb!]
\centering
    \includegraphics[width=1\columnwidth]{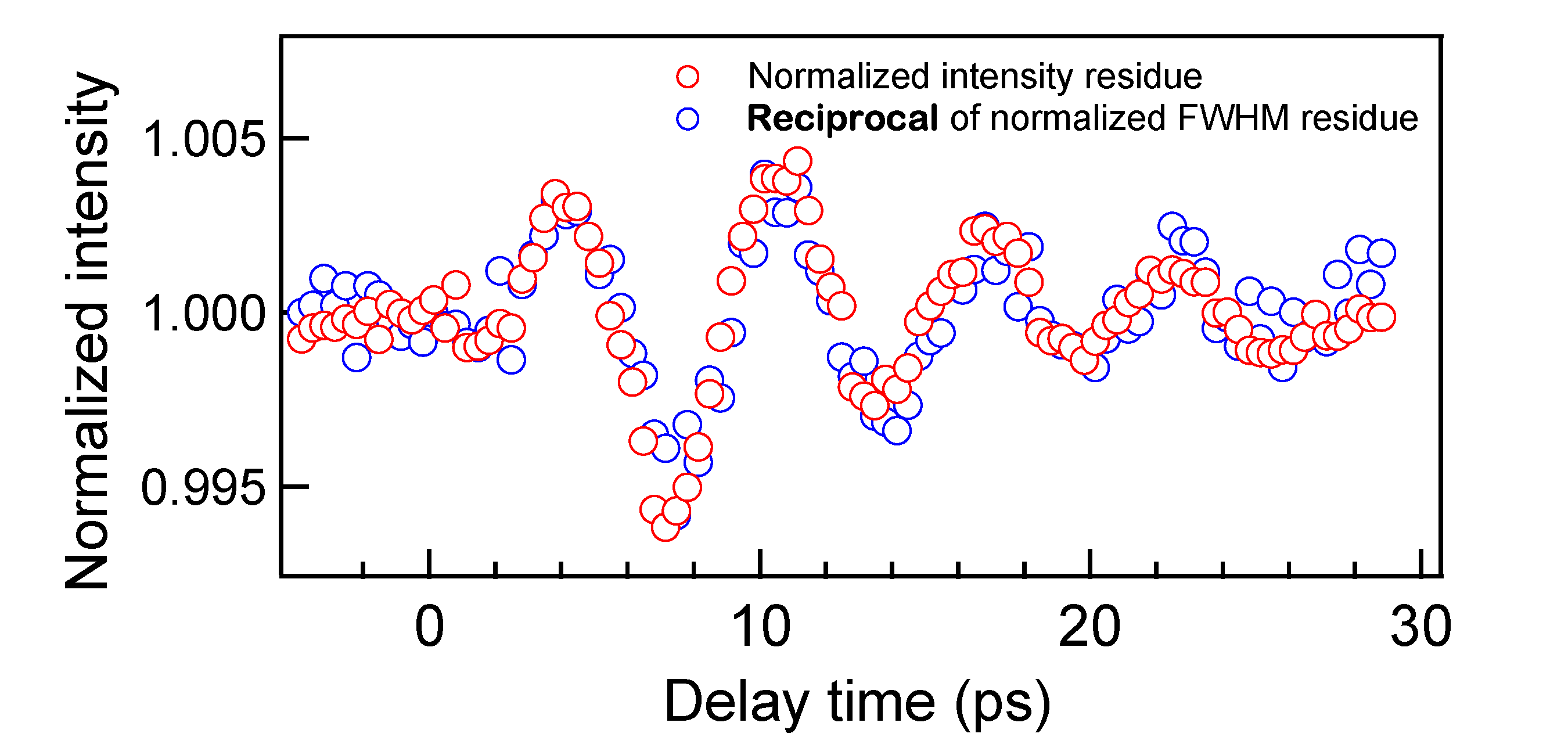}
    \caption{The residue of normalized peak intensity (red) and the \emph{reciprocal} of the residue of peak width (blue) shows a perfect match, which indicates a nearly constant integrated intensity of the Bragg peaks during the wrinkling oscillation.}
\label{fig:reciprocal}
\end{figure}

The above comparison suggests that the integrated intensity measured in each Bragg peak is nearly conserved during the coherent wrinkling dynamics, accounting for the lack of oscillatory signals in its time trace in Fig.~\ref{fig:res}(a). This conservation of integrated intensity also means that the change in the peak intensity must be compensated by a corresponding change in the peak width. To visualize their relation more clearly, we plot the coherent component of normalized peak intensity and the \emph{reciprocal} of its peak width in Fig.~\ref{fig:reciprocal}. The excellent match between the two curves further validates our picture of a coherent wrinkling motion.

\subsection{Approximating tilted rel-rods as cones}

In Fig.~\ref{fig:extract}(d)(e), we sketched the relative positions of the Ewald sphere and the rel-rods in a wrinkled film. We note that Fig.~\ref{fig:extract}(e) is not an exact magnification of Fig.~\ref{fig:extract}(d). Specifically, in Fig.~\ref{fig:extract}(e), all tilted rel-rods intersect with one another at their center and form a cone, which is not the case in the more accurate depiction in Fig.~\ref{fig:extract}(d). Here, we explain why Fig.~\ref{fig:extract}(e) is a reasonable approximation.

\begin{figure}[tb!]
\centering
\includegraphics[width=1\columnwidth]{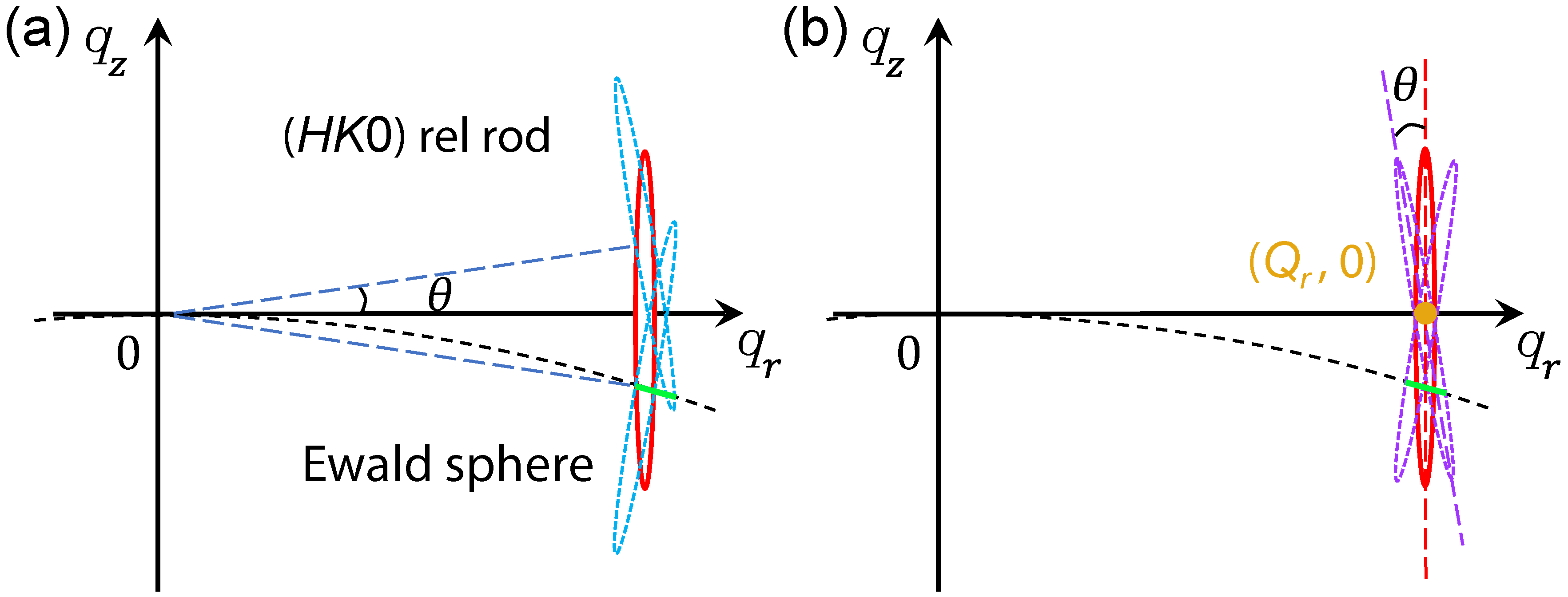}
    \caption{Schematics of the Ewald construction of electron diffraction. The Ewald sphere (solid dashed curve) is modeled by a circle with radius $83.64\text{~{\AA}}^{-1}$ for the 26~keV electron beam. The rel-rods are modeled by narrow ellipses. In order to highlight the minuscule change, the plots are not to scale. (a)~Exact geometry of sample wrinkling in reciprocal space. The blue dashed ellipses represent rotated rel-rods with respect to the origin. (b)~Approximate geometry of sample wrinkling in reciprocal space. The magenta dashed ellipses represent rotated rel-rods with respect to the un-tilted rod center $(Q_r,0)$. The green segments in both panels indicate the width of the intersection, which is proportional to the diffraction peak width.}
\label{fig:geo}
\end{figure}

Without loss of generality, we consider a two-dimensional $q_r$-$q_z$ slice of the momentum space as shown in Fig.~\ref{fig:geo}, where $q_r$ is the in-plane radial momentum transfer within the $(HK0)$ diffraction plane and $q_z$ is the out-of-plane momentum transfer perpendicular to the $(HK0)$ plane. The Ewald sphere has a radius of $R_E=83.64\text{~{\AA}}^{-1}$ for the 26~keV electrons used \cite{Griffiths2013}. Since it passes through the origin, the Ewald sphere satisfies:
\begin{align}
    q_r^2+(q_z+R_E)^2=R_E^2.
\end{align}
For a perfectly flat thin film whose surface normal is parallel to the electron incident direction, we have $Q_r=n\cdot2\pi/a$, where $n$ depends on the diffraction order of the rel-rod. The lattice constant for LCMO is $a = 3.85$~{\AA} in the pseudo-cubic unit cell \cite{Feng2005}. Take the $(020)$ Bragg peak for example, $n=2$, so $Q_r = 3.26\text{~{\AA}}^{-1}$. When the film is tilted, we consider two cases illustrated in Fig.~\ref{fig:geo}: (a)~the rod is rotated by $\pm\theta$ around the origin, and (b)~the rod is rotated around the point $(Q_r,0)$. In case~(a), the major axis of the rotated rod follows: 
\begin{align}
    q_z = \pm k(q_r-Q_r), 
\end{align}
where $k=\arccot\theta$. In case~(b), the major axis follows:
\begin{align}
    q_z\pm Q_r\sin\theta= \pm k(q_r - Q_r\cos\theta).
\end{align}
Adopting the measured wrinkle angle $\theta = 3.71^{\circ}$ in equilibrium (see Sec.~\ref{sec:wrinkle_angle_equilibrium}), we can solve the intersections between the Ewald sphere and the rel-rods, and calculate the peak width for the two geometries in Figs.~\ref{fig:geo}(a) and (b), respectively. Assuming a locally flat Ewald sphere within each rel-rod, we compute the width of intersection, $w$, defined as the distance between the two intersection points (green segments in Fig.~\ref{fig:geo}). In case~(a), $w=0.01118\text{~{\AA}}^{-1}$; in case~(b), $w=0.01092\text{~{\AA}}^{-1}$. The difference in $w$ between these two cases is on the order of $10^{-4}\text{~{\AA}}^{-1}$, which is negligible compared to the peak width $w$ itself as well as the length of the rel-rod (0.03~{\AA}$^{-1}$). Therefore, we can use Fig.~\ref{fig:geo}(b) to approximate the rel-rod intersection geometry and assume that different rel-rods from a wrinkled film form a cone-like geometry for peaks with low diffraction orders.

\begin{figure*}[tbh!]
\centering
\includegraphics[width=0.75\textwidth]{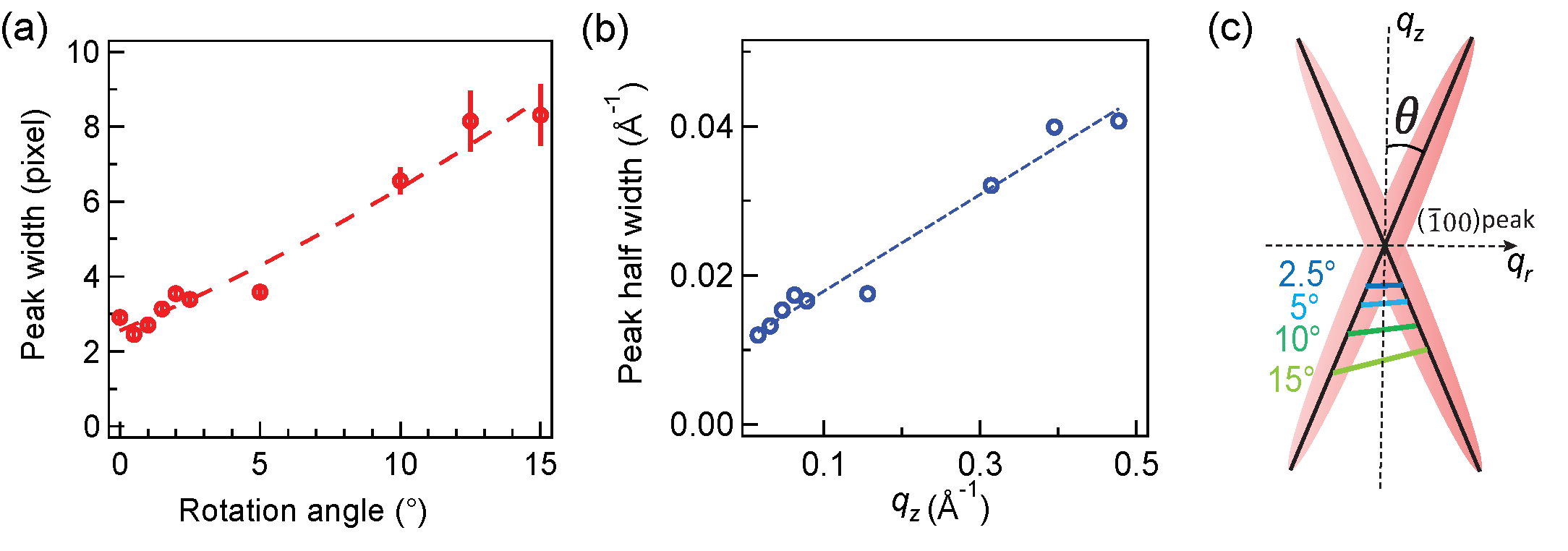}
    \caption{(a)~Peak width measured by FWHM of the $(\overline{1}00)$ peak as the sample is rotated around $b$-axis. (b)~Projected half-width at half maximum of the peak as a function of $q_z$ [see schematic in Fig.~\ref{fig:geo}(b)]. (c)~Schematic of a momentum-space cone for a wrinkled film with wrinkle angle $\theta$. Colored curves denote segments of the Ewald sphere corresponding to different rotation angles.}
\label{fig:slice}
\end{figure*}

\subsection{Extraction of wrinkle angle in equilibrium\label{sec:wrinkle_angle_equilibrium}}

To measure the wrinkle angle $\theta$ in equilibrium, we can rotate the Ewald sphere to intersect the cone at different $q_z$ (see schematic in Fig.~\ref{fig:geo}). For a fixed electron beam incidence, we can equivalently rotate the sample, following a similar procedure in Ref.~\cite{Kumar2015} [see Fig.~\ref{fig:slice}(c)]. We rotate the crystal around the $b$-axis and measure the width of the $(\overline{1}00)$ peak, shown in Fig.~\ref{fig:slice}(a). By converting the rotation angle to $q_z$ following the schematic in Fig.~\ref{fig:geo}(b), we obtain the relation between the peak width and $q_z$, shown in Fig.~\ref{fig:slice}(b). If we denote the line slope in Fig.~\ref{fig:slice}(b) as $s$, the opening angle of the cone, namely the wrinkle angle, is $\theta = \arctan s = 3.71\pm0.25^\circ$. 

\subsection{Projected reciprocal lattice constant change\label{sec:projected_reciprocal_lattice_constant_change}}

As indicated in Fig.~\ref{fig:extract}(e), when a cone of rel-rods intersects with the Ewald sphere, the higher intensity concentrates towards the (000) beam. This slightly skewed intensity distribution causes the apparent reciprocal lattice constant to decrease when the cone broadens, leading to a smaller projected BZ area as discussed in the main text. Here, we quantify this effect and establish the relation between a change in peak width and a change in the pBZ area.

\begin{figure}[th]
\centering
\includegraphics[width=0.6\columnwidth]{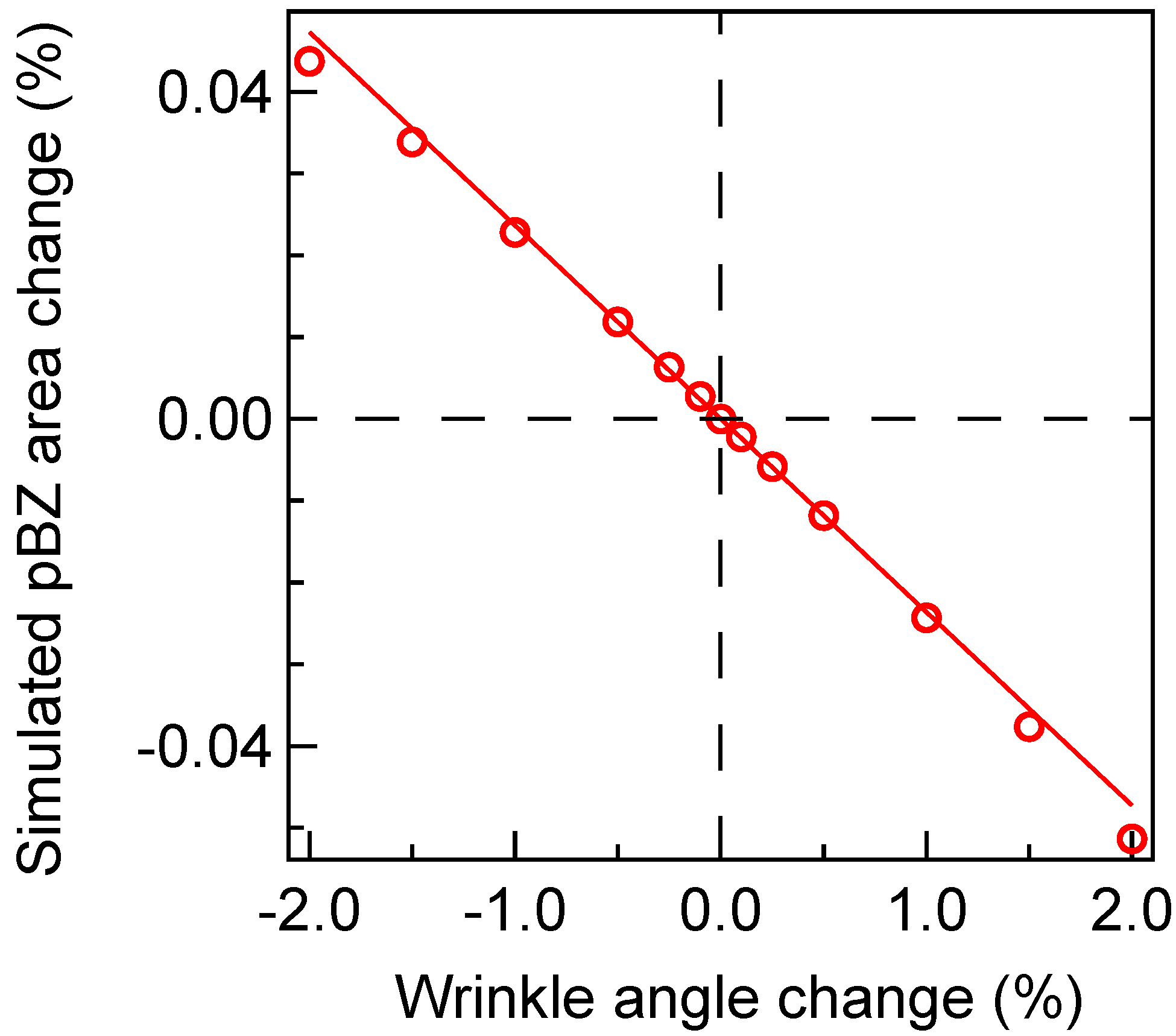}
    \caption{Percentage change in the simulated pBZ area as a function of the wrinkle angle change $\Delta\theta/\theta_0$. Solid line is a linear fit. Within $\pm 1\%$ change of the equilibrium wrinkle angle $\theta_0 = 3.71^\circ$, the trend is quasi-linear, giving the $-0.026$ scaling factor between the pBZ area change and the wrinkle angle change.}
\label{fig:simVangle}
\end{figure}

We again follow the Ewald sphere construction in Fig.~\ref{fig:geo}(b). As mentioned in Sec.~\ref{sec:integrated_intensity}, we assume a Gaussian distribution of diffraction intensity $I(Q_z)$ along the major axis of the rod parameterized by $Q_z$, where $Q_z = 0$ marks the center of the rel-rod. The Gaussian function has a characteristic width of $2\pi/T$, where $T=200$~{\AA} is the film thickness. Given a particular wrinkle angle $\theta$, we compute the intensity distribution along the circumference of the Ewald ``circle'' [dashed curve in Fig.~\ref{fig:geo}(b)], assuming that the local tilt angle $\theta_\text{local}$ of a particular rod within the cone follows a Gaussian distribution centered at $\theta_\text{local} = 0$. Given this intensity distribution, we compute its center of mass, which is then projected to the detector plane to precisely determine the diffraction peak position. This calculation was performed for each peak in our diffraction pattern, and the positions of all peaks allow us to extract the simulated pBZ area through the global fitting procedure described in Sec.~\ref{sec:BZ_area_fitting}. By computing the simulated pBZ area with a range of wrinkle angle $\theta$ around its equilibrium value of 3.71$^\circ$ (Fig.~\ref{fig:simVangle}), we find that a 1\% increase in the peak width  [Eq.~\eqref{eq:theta_FWHM}] corresponds to a 0.026\% decrease in the pBZ area.

\begin{figure}[th]
\centering
\includegraphics[width=0.8\columnwidth]{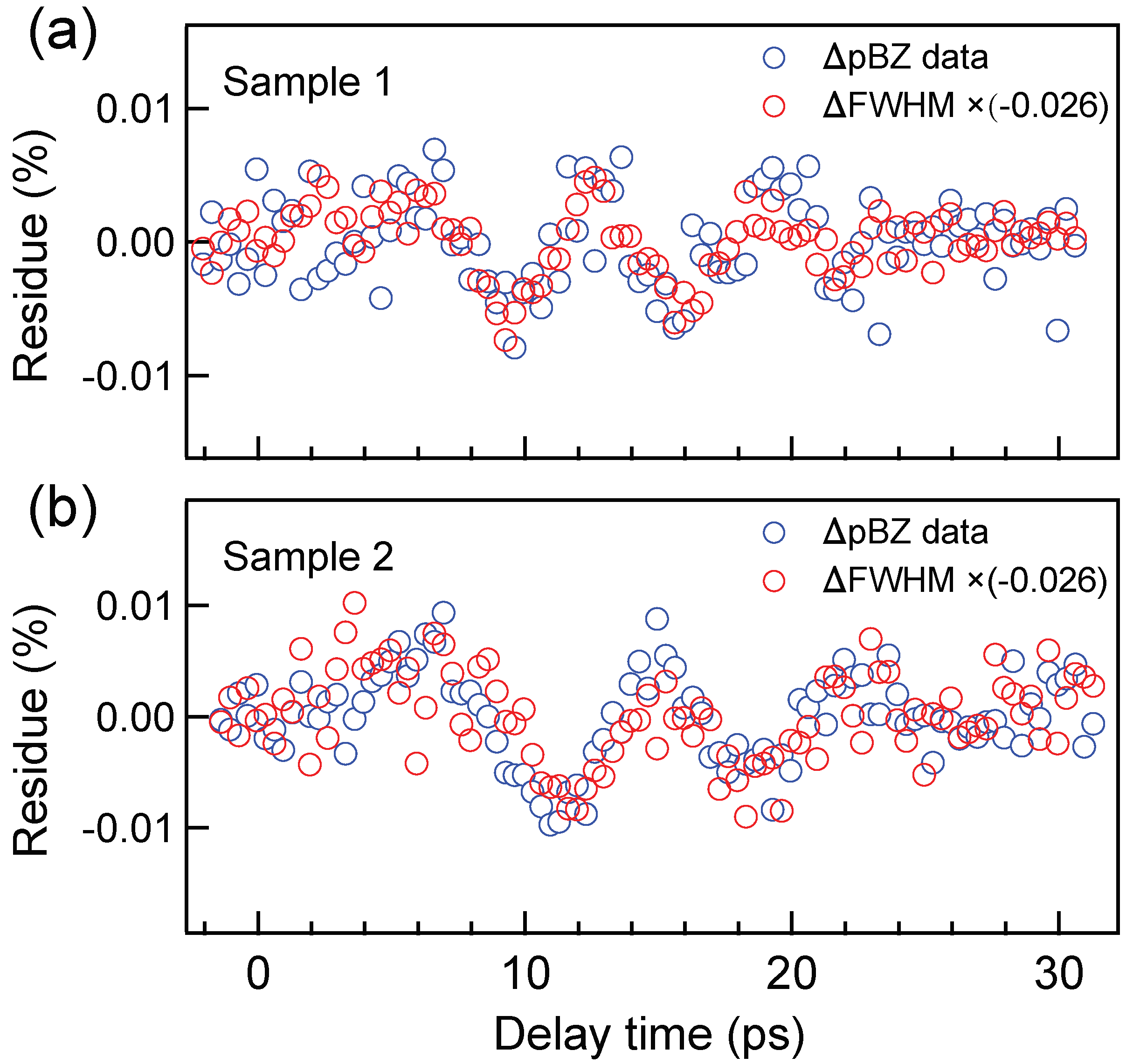}
    \caption{\textit{Blue}:~Oscillatory signal of the pBZ area change measured from Sample~1 (a) and Sample~2 (b). These two samples share the same thickness but are mounted on different TEM grids. \textit{Red}:~Oscillatory signal of the peak width change multiplied by the calculated scaling factor.}
\label{fig:consistency}
\end{figure}

The calculation above only depends on the electron beam energy, the lattice constant, and the film thickness. Thus, the same relation between the pBZ area and peak width should hold for different LCMO samples with the same thickness under the same electron beam condition. To demonstrate the validity of the above calculation, Fig.~\ref{fig:consistency} show the data obtained with an incident fluence of 2~mJ/cm$^2$, half the value applied for the data in Fig.~\ref{fig:res} in the main text. Here, Fig.~\ref{fig:consistency}(a) is measured from the same sample as in the main text while Fig.~\ref{fig:consistency}(b) is acquired another sample. In both cases, the calculated scaling factor of $-0.026$ between the peak width change (red) and the pBZ area change (blue) shows an excellent agreement with the experimental data.

\begin{figure}[thb!]
\centering
\includegraphics[width=1\columnwidth]{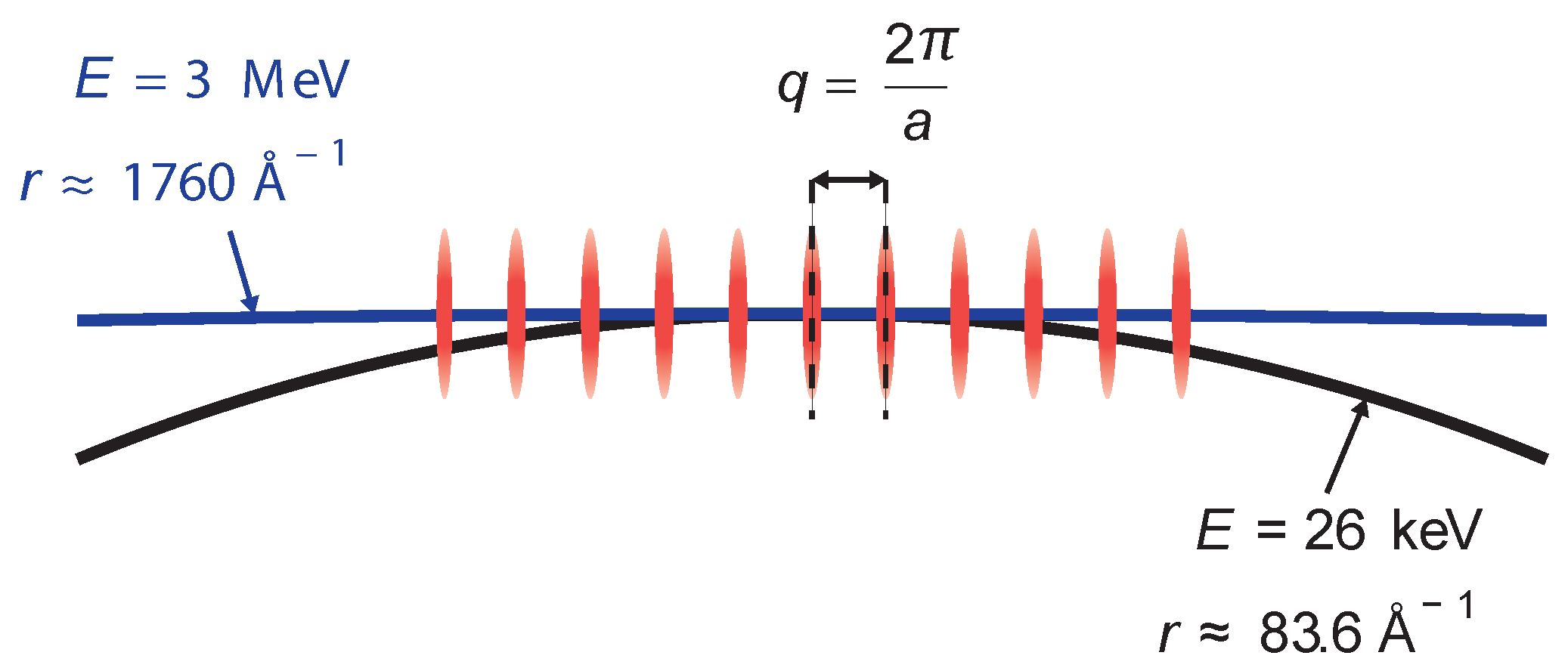}
    \caption{The Ewald spheres of MeV (blue) and keV (black) electron diffraction, plotted to scale. The size of rel-rods (red ellipses) is exaggerated for better visibility. The values of respective electron kinetic energies are labeled, and $r$ denotes the radius of the Ewald sphere. The curvature of the MeV Ewald sphere is negligible within the momentum range of a typical electron diffraction measurement.}
\label{fig:keV_vs_MeV}
\end{figure}

In addition, we note that for MeV ultrafast electron diffraction experiments, such as the one performed in Ref.~\cite{Mannebach2015}, the Ewald sphere is nearly flat close to the undiffracted beam. In this case, shift in the diffraction peak position caused by width broadening is negligible: Following the same calculation, for a 1\% change width change, the position change is within the numerical accuracy ($<0.00002$\%). To help visualize the comparison between keV and MeV electron diffraction,  Fig.~\ref{fig:keV_vs_MeV} shows the Ewald spheres of 26~keV (this work) and 3~MeV (SLAC) electron beams, where the latter can essentially be treated as flat. Therefore, the peak position change in Ref.~\cite{Mannebach2015} was interpreted as a change in the lattice parameter and hence strain. On the other hand, in the present work of LCMO, the peak position oscillation has minimal contribution from lattice strain, as detailed in the next section.

\section{P\lowercase{rojected} B\lowercase{rillouin zone area change under fixed and free boundary conditions}\label{sec:boundary_condition}}

Based on a comparative study of two boundary conditions in Fig.~\ref{fig:bdc}, we concluded in the main text that the contact points between the thin film and the copper grid are not fixed. This conclusion relies on the agreement between the measured pBZ area change and the calculated pBZ area based on the wrinkle angle oscillation for a free but not a fixed boundary condition. In this section, we explain the calculation in more details. 

\begin{figure}[thb!]
\centering
\includegraphics[width=0.9\columnwidth]{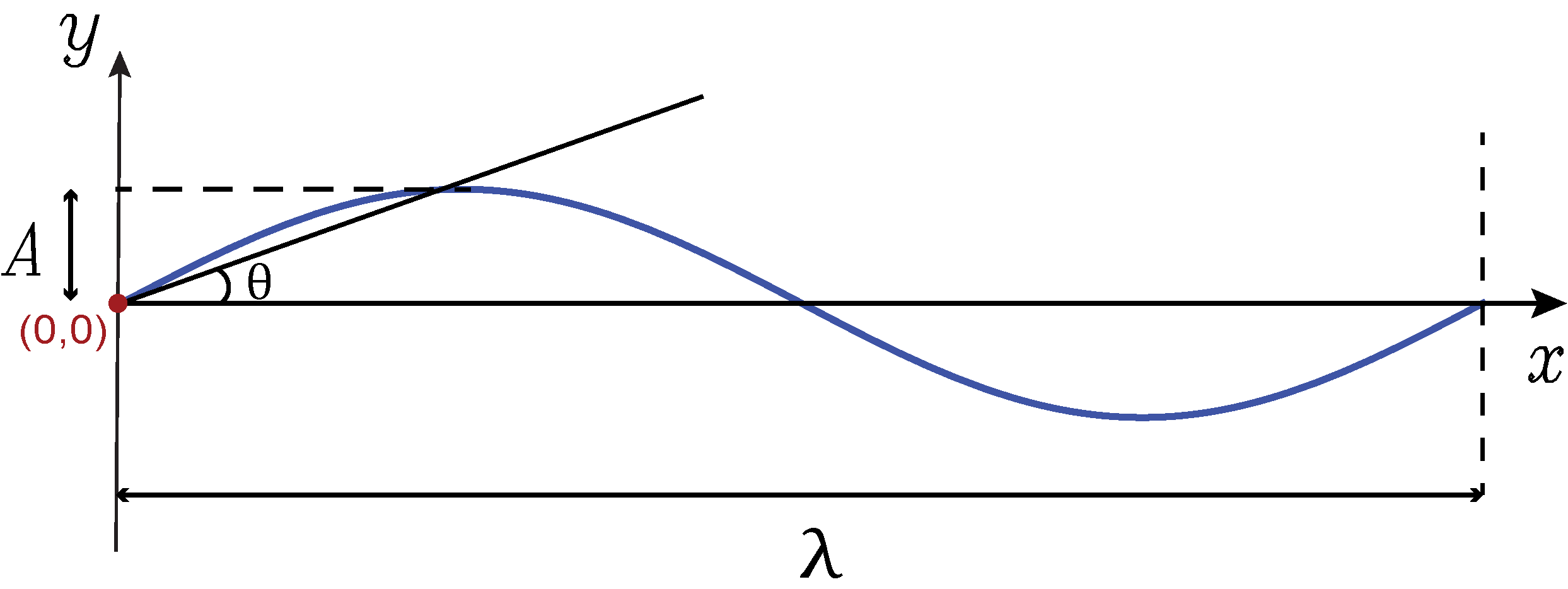}
    \caption{Cross-sectional view of a wrinkled film described by a sinusoidal path.}
\label{fig:sine}
\end{figure}

For a free boundary condition, the dominant factor that contributes to the measured pBZ area change is the asymmetric intersection between rel-rod cones and the Ewald sphere, as detailed in Sec.~\ref{sec:projected_reciprocal_lattice_constant_change}. In this case, the scaling factor between the pBZ area change and the peak width change is $-0.026$, which well captures the $\sim50$ times difference between the oscillation amplitudes of these two quantities.

For a fixed boundary condition, if the thin film is initially strain-free, as the wrinkle angle $\theta$ increases, the film will become more strained. Without loss of generality, we consider a simplified model of a sinusoidal path to capture the wrinkle and strain dynamics of the film \cite{Du2021}. As shown in Fig.~\ref{fig:sine}, the cross-sectional view of a wrinkled film is described in Cartesian coordinates by
\begin{align}
    y(x)= A\sin\left(\frac{2\pi x}{\lambda}\right),
\end{align}
where $A$ is the amplitude of the out-of-plane modulation and $\lambda$ is an effective wavelength of the wrinkle. We estimated $\lambda$ from the optical microscopy image in Fig.~\ref{fig:main}(a). By counting the number of full-wavelength wrinkles in each of the 70$\times$70~$\upmu$m$^2$ mesh grids and taking the average, we find $\lambda\approx22~\upmu$m. From the geometry in  Fig.~\ref{fig:sine}, we further have
\begin{align}
    A=\frac{\lambda}{4}\tan\theta,\label{eq:A_theta}   
\end{align}
where $\theta$ is taken as the wrinkle angle. We then solve for the path length of the sinusoid given $\theta$ and $\lambda$. The percentage change of the path length is the lattice strain, with which we can calculate the changes in the reciprocal lattice constant and the pBZ area. As shown in Fig.~\ref{fig:bdc}(b), the calculated pBZ area change is an order of magnitude larger than the measured change, hence invalidating the assumption of a fixed boundary condition that forms the premise of this calculation.

\section{S\lowercase{train dynamics of the wrinkling film}\label{sec:strain_dynamics}}

\begin{figure}[tb!]
\centering
\includegraphics[width=0.4\textwidth]{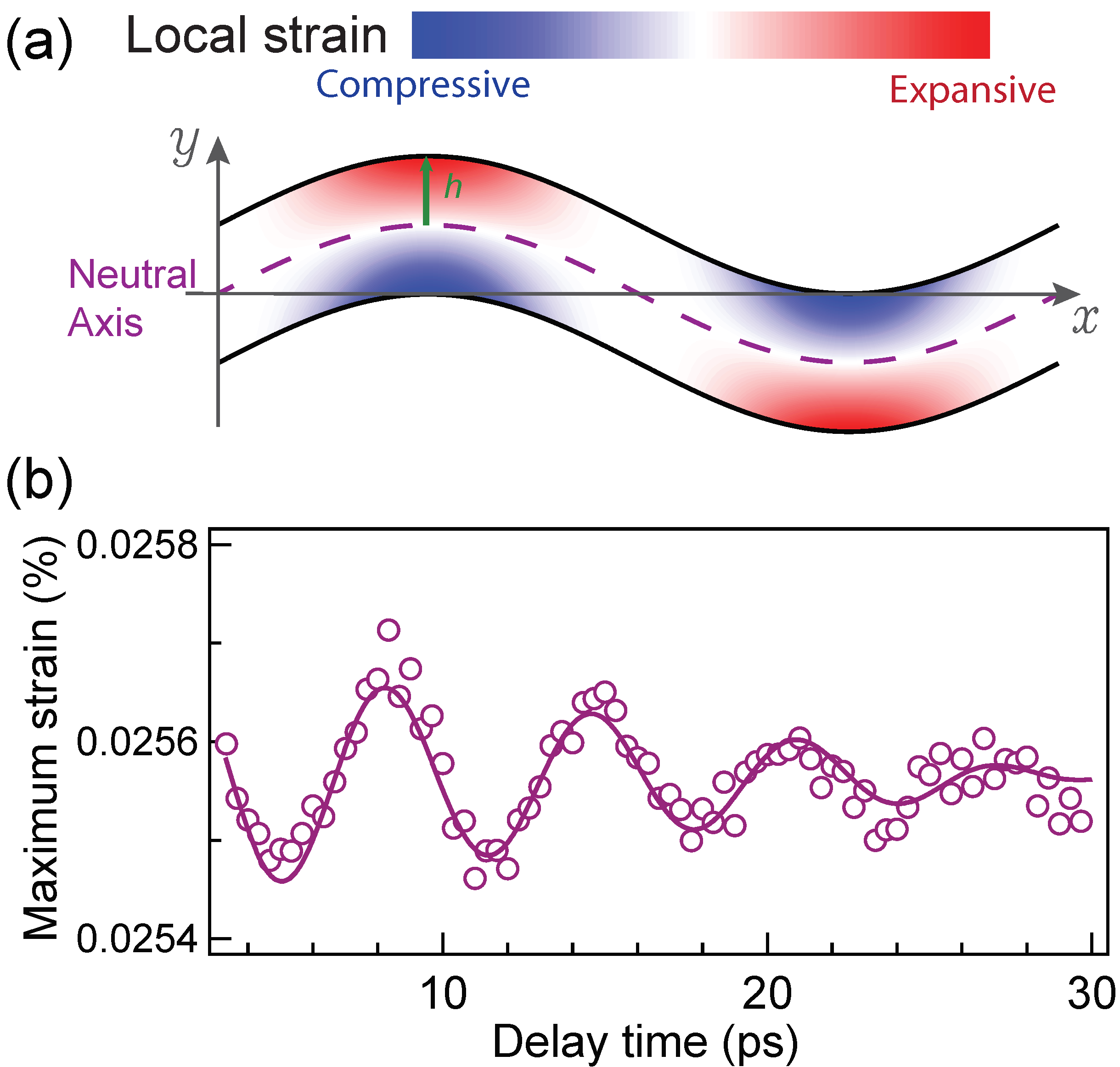}
    \caption{(a)~Schematic of local strains in a wrinkled film. The absolute values of compressive and expansive strains are the same by symmetry. (b)~Time evolution of the maximum local strain calculated based on the wrinkle angle dynamics in Fig.~\ref{fig:extract}(f).}
\label{fig:strain}
\end{figure}

With a free boundary condition demonstrated in Sec.~\ref{sec:boundary_condition}, we argued that there is a negligible lattice constant change due to wrinkle formation and hence a very small change in the pBZ area. Nonetheless, along the thickness direction of the film, denoted by $h$, a local compressive or expansive strain can build up where the curvature of the film is large [Fig.~\ref{fig:strain}(a)]. Here we provide an estimate of the maximum strain in the LCMO thin films following the same sinusoidal model discussed in Sec.~\ref{sec:boundary_condition}, with the additional consideration of the film thickness.

We follow the same methodology in Ref.~\cite{Du2021}, as shown in Fig.~\ref{fig:strain}(a). The bending strain as a function of $x$ and $h$ is given by
\begin{align}
    \varepsilon = \frac{h}{R(x)},\label{eq:strain-def}
\end{align}
where the local radius of curvature defined with respect to the neutral axis is [Fig.~\ref{fig:strain}(a)]:
\begin{align}
R(x)=\frac{\left[1+(y'(x))^{2}\right]^{3/2}}{y''(x)}.
\end{align}
The film thickness is 20~nm, so $h\in[-10,10]$~nm. The maximum strain occurs at $x=\lambda/4$ and $h=10$~nm, where the radius of curvature is
\begin{align}
    R(\lambda)=\frac{\lambda^2}{(2\pi)^2A}.
\end{align}
Based on Eq.~\eqref{eq:A_theta}, we re-write $R(\lambda)$ as
\begin{align}
    R=\frac{\lambda}{\pi^2\theta},\label{eq:lambda_theta}
\end{align}
where we assume $\theta\approx\tan\theta$ for a small wrinkle angle $\theta$. Given the free boundary condition, $\lambda$ changes with the wrinkle angle $\theta$ while the path length of the sinusoid within one period remains constant. We thus have an approximate relation between $\lambda$ and $\theta$ (see Fig.~\ref{fig:sine}):
\begin{align}
\frac{\lambda(t)}{\cos\theta(t)}= \frac{\lambda_0}{\cos\theta_0},\label{eq:lambda_cos_theta}
\end{align}
where $\lambda_0$ and $\theta_0$ are values in equilibrium and $t$ denotes the delay time. Substituting Eq.~\eqref{eq:lambda_cos_theta} into Eq.~\eqref{eq:lambda_theta} yields:
\begin{align}
    R(t) = \frac{\lambda_0\cos\theta(t)}{\pi^2\theta(t)\cos\theta_0},\label{eq:R_time}
\end{align}
where $\lambda_0 \approx 22~\upmu$m (Sec.~\ref{sec:boundary_condition}) and $\theta_0 =3.71^\circ$ (Sec.~\ref{sec:wrinkle_angle_equilibrium}). From Eqs.~\eqref{eq:strain-def} and \eqref{eq:R_time}, we establish the relation between wrinkle-induced maximum strain $\varepsilon_\text{max}$ and the time-dependent wrinkle angle $\theta(t)$:
\begin{align}
    \varepsilon_\text{max}[\theta(t)] = \frac{\pi^2\theta(t)\cos\theta_0\cdot10~\text{nm}}{22~\upmu\text{m}\cdot\cos\theta(t)}.
\end{align}
This equation allows us to compute the time evolution of the maximum strain in a wrinkled film based on the oscillation of the wrinkle angle $\theta$; the result is shown in Fig.~\ref{fig:strain}(b).  

\begin{figure}[tb!]
\centering
\includegraphics[width=0.8\columnwidth]{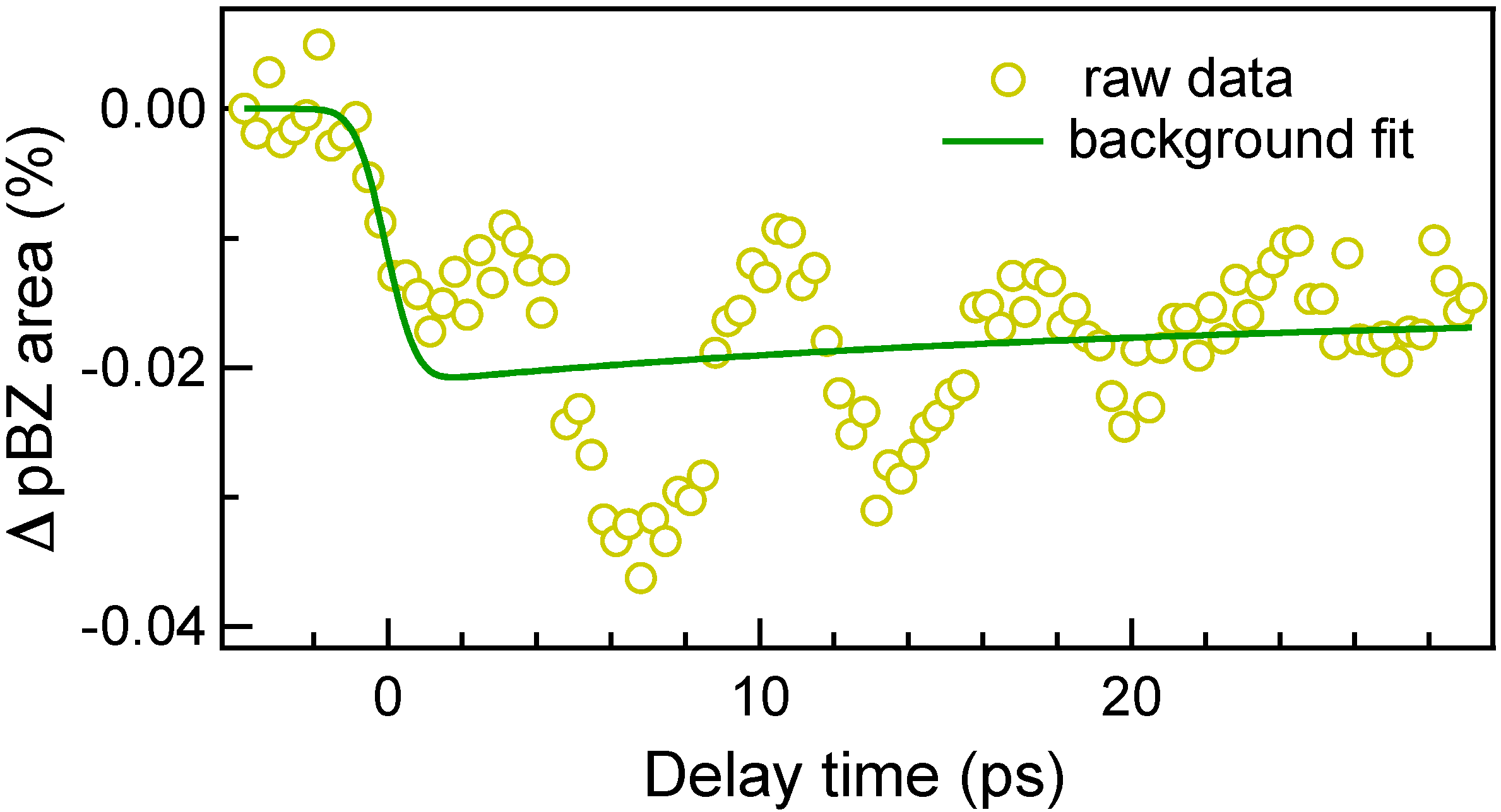}
    \caption{Time evolution of the pBZ area, reproduced from Fig.~\ref{fig:main}(f), with a fitted curve of the incoherent response according to Eq.~\eqref{eq:fit}. The overall decrease in the pBZ area is caused by a thermoelastic expansion following photoexcitation where the lattice constant increases by approximately 0.01\%.}
\label{fig:thermal_expansion}
\end{figure}

Let us compare the peak-to-peak change in the maximum local strain during the wrinkling oscillation in Fig.~\ref{fig:strain}(b) to the overall lattice strain induced by thermal expansion after photoexcitation. The thermal expansion is a result of the energy transfer from excited hot electrons to the lattice, and is thus an incoherent process. The expansion will lead to an overall decrease of the projected pBZ area over the same $\sim2$~ps timescale as in the integrated intensity [Fig.~\ref{fig:res}(a)], which is best illustrated in Fig.~\ref{fig:thermal_expansion} where we fitted the pBZ area response to Eq.~\eqref{eq:fit} in order to quantify its initial decrease. While one can alternatively interpret this incoherent decrease of the pBZ area as a manifestation of an overall increase in the wrinkle angle, such a scenario is incongruent with the change in the peak width after photoexcitation [Fig.~\ref{fig:main}(e)], which does not show an overall increase in its time-averaged value. From the overall pBZ area decrease in Fig.~\ref{fig:thermal_expansion}, we estimate the thermal expansion-induced strain to be 0.01\%, which is 50~times larger than the peak-to-peak variation of the maximum local strain during the wrinkling oscillation [Fig.~\ref{fig:strain}(b)]. This comparison further illustrates that the ultrafast wrinkle formation in LCMO is free from any significant strain build-up, and the periodic delamination at the film-substrate contact points is key to the strain relief mechanism.

\section{E\lowercase{stimation of the delamination length}\label{sec:strain_dynamics}}

In the main text, we stated that the boundary of the thin film moves by approximately 2~nm between the equilibrium position and the maximum transient wrinkle angle. We obtained this value by considering that the boundary movement is a result of a changing area of the freestanding part of the film due to strain relief. In the sinusoidal model of a wrinkled film, the boundary displacement equals to the length difference in the suspended arcs between the copper bars [Fig.~\ref{fig:bdc}(b)]. Following the assumption of Eq.~\eqref{eq:lambda_cos_theta}, the arc length within a single period of the sinusoid is constant, but the periodicity $\lambda$ decreases when the wrinkle angle increases. Let $n$ denote the number of periods between the two copper bars, where $n$ is not necessarily an integer. Then the length difference is approximately $n|\lambda(t)-\lambda_0|$, yielding 4~nm between the maximum wrinkle angle and the equilibrium position during the photoinduced oscillation. As there are two ends of the film, the displacement in each boundary is $\sim2$~nm. As the frequency of the wrinkling oscillation is 0.16~THz, this displacement over a quarter of the oscillation period corresponds to an average delamination speed of 1~km/s. For reference, the propagation speed of the acoustic wave corresponding to the photoinduced oscillation in the 20-nm-thick film is $2\cdot20\,\text{nm}\cdot0.16\,\text{THz}=6.4$~km/s, which is more than six times higher than the delamination speed.

\end{document}